\documentclass[cernpreprint,atlasdraft=true,coverpage=false,texlive=2016,UKenglish,texmf]{atlasdoc}
 
\usepackage{atlaspackage}
\usepackage{mathrsfs}
\usepackage{atlasbiblatex}
 
\usepackage{atlasphysics}
 
\graphicspath{{logos/}{figures/}}
 
\usepackage{ANA-EXOT-2018-63-PAPER-defs}

\AtlasTitle{Search for dark matter in association with an energetic photon in $pp$ collisions at $\sqrt{s} = 13$~\ensuremath{\text{Te\kern -0.1em V}} with the ATLAS detector}

\PreprintIdNumber{CERN-EP-2020-178}
 
\AtlasJournalRef{JHEP 02 (2021) 226}
\AtlasDOI{10.1007/JHEP02(2021)226}

\AtlasAbstract{
A search for dark matter is conducted in final states containing a photon and missing transverse momentum  in proton--proton collisions at $\sqrt{s} = 13$~\ensuremath{\text{Te\kern -0.1em V}}.
The data, collected during 2015--2018 by the ATLAS experiment at the CERN LHC, correspond to an integrated luminosity of 139~$\mathrm{fb}^{-1}$.
No deviations from the predictions of the Standard Model are observed and 95\% confidence-level upper limits between 2.45~fb and 0.5~fb  are set on the visible cross section for contributions from physics beyond the Standard Model, in different ranges of the missing transverse momentum.
The results are interpreted as 95\% confidence-level limits in models where weakly interacting dark-matter candidates are pair-produced via an $s$-channel axial-vector or vector mediator.
Dark-matter candidates with masses up to 415 (580)~\ensuremath{\text{Ge\kern -0.1em V}} are excluded for axial-vector (vector) mediators, while the maximum excluded mass of the mediator is 1460 (1470)~\ensuremath{\text{Ge\kern -0.1em V}}.
In addition, the results are expressed in terms of 95\% confidence-level limits on the parameters of a model with an axion-like particle produced in association with a photon, and are used to  constrain the coupling $g_{aZ\gamma}$ of an axion-like particle to the electroweak gauge bosons.}

\author{The ATLAS Collaboration}
 
\AtlasRefCode{EXOT-2018-63}
\AtlasDate{\today}

\hypersetup{pdftitle={ATLAS document},pdfauthor={The ATLAS Collaboration}}
 
\begin{document}
 
\maketitle
 
\tableofcontents
 
\section{Introduction}
Despite its astounding success, the Standard Model (SM) of particle physics is considered to be a low-energy approximation to some 
fundamental  theory of nature, with new degrees of freedom and symmetries that would be evident at a higher energy.
Numerous attempts have been made to figure out a footprint of physics beyond the Standard Model (BSM) at the Large Hadron Collider (LHC), but no significant sign of new physics has been observed yet.
In this paper, proton--proton collisions at the LHC are used to explore the production of events with an energetic photon and large missing transverse momentum (\metv, with magnitude \met), as it may constitute a striking signature of BSM physics, including extensions that account for particle dark matter (DM)~\cite{Clowe_2006}.
 
Understanding the nature of dark matter has provided some of the strongest motivations to search for BSM physics at the LHC.
Recent advances in theory and experiment combining high-energy physics, astrophysics and cosmology have led to a potential DM phenomenology with a variety of experimental signatures that could be observed in proton--proton collisions at the LHC.
Among the searches are those focusing on final states that try to replicate the annihilation processes which could have led to dark-matter freeze-out in the early universe.  This is the case for direct DM pair-production yielding large \met\ in association with a visible particle (X) that, in most searches, originates from initial-state radiation~\cite{Goodman_2010}.
The $\textrm{X} + \met$ signature is the hallmark of this type of search. 
 
Axion-like particles (ALPs), originally motivated by the QCD axion from the dynamical solution to the strong CP problem of the SM~\cite{PhysRevLett.38.1440}, provide a compelling and elegant explanation of DM. More generally, ALPs appear in any theory with a spontaneously broken global symmetry.  They could be non-thermal DM candidates~\cite{Preskill1983} or mediators to a dark sector.  Depending on the different ALP production mechanisms at the LHC, it is possible to probe a large range of masses and couplings over several orders of magnitude~\cite{Bauer_2019}. Masses from \keV\ to \MeV\ are conducive to a stable ALP and are best searched for in association with either a photon or a jet, and hence the expected final states are events with a photon or a jet plus  \met~\cite{Jaeckel_2013, Mimasu2015, Brivio:2017ije}.

This paper presents the results of a search for an excess of events with a $\gam + \met$ final state over the SM prediction.
The search is performed using the full Run-2 dataset collected by the ATLAS experiment at the LHC in proton--proton collisions at a centre-of-mass energy of $\sqrt{s} = 13~\TeV$.
The $\gam + \met$ final state has the advantage of a clean signature that nicely complements the other $\textrm{X} +  \met$ processes.  The sensitivity of this search is enhanced compared to a previous search using 36.1~\ifb~\cite{paper_monophoton2017} due to the increased size of the dataset and by incorporating new criteria for reconstructed objects. The $\textrm{X} + \met$ signatures have been explored by the LHC experiments for the cases where X is a photon~\cite{paper_monophoton2017,Sirunyan:2018dsf}, a jet~\cite{Aaboud_2018JHEP126, PhysRevD.97.092005}, a heavy quark~\cite{Aaboud2018JHEP18, Sirunyan_2018}, a vector boson~\cite{ PhysRevD.97.092005,CMSfix,Aaboud:2018xdl} or a Higgs boson~\cite{Aaboud_2017_PhysRevD.96.112004, Sirunyan_2017_JHEP10}.
 
The results are interpreted using simplified models where Dirac-fermion DM candidates (denoted by $\chi$) interact with quarks through the exchange of a mediator in the $s$-channel via vector or axial-vector interactions~\cite{Abdallah, Buchmueller, dmforum}.
A photon can be radiated from the initial state and the $\chi\bar{\chi}$ pair is invisible to the detector, resulting in a $\gam + \met$ final state as shown in Figure~\ref{fig:Feynman diagrams} (left).
 
\begin{figure}[ht!]\centering
\includegraphics[width=0.33\textwidth]{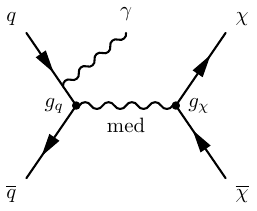}
\includegraphics[width=0.33\textwidth]{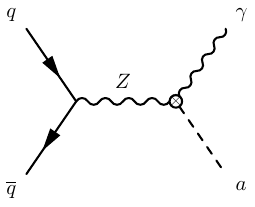}\\
\caption{Feynman diagrams   corresponding to the simplified DM  model  (left) and ALP DM  model (right) considered.}
\label{fig:Feynman diagrams}
\end{figure}
 
The  free parameters in models of this kind are the mass of the mediator, $m_{\mathrm{med}}$, the mass of the DM particle, $m_\chi$,
the couplings $g_q$, $g_\ell$ and $g_\chi$ of the mediator to quarks, leptons and DM particles respectively, and the width of the mediator, $\Gamma_{\textrm{med}}$~\cite{DMwg}.
The latter is computed as the minimum width allowed given the couplings and masses.

In addition, a model with an ALP (denoted by  $a$) produced  in association with a photon is used. The model considered is an effective field theory (EFT) that extends the SM Lagrangian, by using effective operators to describe the interactions of ordinary matter with an additional particle. This particle is a generic CP-odd (pseudo-)Nambu--Goldstone boson of a spontaneously broken symmetry at energies below a scale higher than the electroweak scale, singlet under the SM charges and playing the role of the ALP~\cite{Brivio:2017ije}.

Signals from this model consist of events having the ALP generated in association to a photon as shown in Figure~\ref{fig:Feynman diagrams} (right).  In this ALP EFT, the new physics scale to be considered is the ALP decay constant  $f_a$, which 
regulates the dimension-5 operators built from the SM fields and  $a$. After electroweak
symmetry breaking,  the couplings of ALPs to the electroweak gauge bosons can be obtained  as a linear combination of the relevant free parameters, namely the real operator coefficients  $c_i$ in the effective Lagrangian describing bosonic ALP couplings, and  the effective scale $f_a$. The ALP mass is supposed to be a free parameter, but since the ALP is considered a light particle ($\sim 1~\MeV$) it does not affect the kinematics. The coupling of ALPs to two photons is tightly constrained by experimental observations, and it is taken to be zero, which allows a further reduction in the number of parameters (see Ref.~\cite{Brivio:2017ije} and the references therein). 
The resulting limits for this model are set on  the energy scale $f_a$ and $\CW$, the coefficient for the operator built from the $SU(2)_L$ gauge group field and $a$.
The signal cross section depends on the square of the ratio of an operator coefficient to the effective scale, $(c_i/f_a)^2$, and hence the ratio $\CW$/$f_a$ is the relevant combination of parameters provided by  this analysis.

 
\section{ATLAS detector}
 
The ATLAS experiment is a multipurpose detector~\cite{PERF-2007-01, ATLAS-TDR-19,PIX-2018-001} with a cylindrical geometry and almost 4$\pi$ coverage in solid angle.\footnote{ATLAS uses a right-handed coordinate system with its origin at the nominal  interaction point (IP) in the centre of the detector and the $z$-axis along the beam pipe. The $x$-axis points from the IP to the centre of the LHC ring, and the $y$-axis points upward. Cylindrical coordinates ($r,\phi$) are used in the transverse plane, $\phi$ being the azimuthal angle around the beam pipe. The pseudorapidity is defined in terms of the polar $\theta$ angle as ${\eta = -\ln\left[\tan(\theta/2)\right]}$.}
The collision point is surrounded by tracking detectors, collectively referred to as the inner detector (ID), followed by a superconducting solenoid providing a 2~T axial magnetic field, a calorimeter system and a muon spectrometer.
 
The ID provides precise measurements of charged-particle tracks in the pseudorapidity range $|\eta| <$ 2.5. It consists of three subdetectors arranged in a coaxial geometry around the beam axis: a silicon pixel detector, a silicon microstrip detector and a transition radiation tracker.
 
The electromagnetic (EM) calorimeter covers the region $|\eta|< 3.2$ and is based on a high-granularity, lead/liquid-argon (LAr) sampling technology.
It is segmented longitudinally in shower depth. The first layer has a high granularity in the $\eta$ direction in order to provide efficient discrimination between single-photon showers and two overlapping photons originating from a $\pi^0$ decay.
The second layer is where most of the energy, deposited in the calorimeter by electron- or photon-initiated electromagnetic showers, is collected.
Significant energy deposits can be left in the third layer by very high energy showers; this layer can also be used to correct for energy leakage beyond the
electromagnetic calorimeter.
 
The hadronic calorimeter uses a steel/scintillator-tile sampling detector in the region $|\eta| < 1.7$ and a copper/LAr detector in the region 1.5~$<|\eta|<$~3.2. The forward calorimeter (FCAL) covers the range 3.2~$<|\eta|<$~4.9 and uses LAr as the active material and copper or tungsten as absorbers for the EM and hadronic sections, respectively.
 
The muon spectrometer (MS) consists of separate trigger and high-precision tracking chambers to measure the deflection of muons in a magnetic field generated by three large superconducting toroids arranged with an eightfold azimuthal coil symmetry around the calorimeters. The high-precision chambers cover the range of  $|\eta| < 2.7$. The muon trigger system covers the range $|\eta| < 2.4$ with resistive-plate chambers in the barrel and thin-gap chambers in the endcap regions.
 
A two-level trigger system is used to select events in real time~\cite{Trigger}. It consists of a hardware-based first-level trigger and a software-based high-level trigger. The latter employs algorithms
similar  to those used in the offline reconstruction.

 
\section{Dataset and simulated events}
\label{sec:MCSimulations}
 
The analysis is performed on a set of proton-proton collision data collected by the ATLAS detector at $\sqrt{s}$~=~13~\TeV\ between 2015 and 2018. In this period, the LHC delivered colliding beams with a peak instantaneous luminosity up to $L = 2.1 \times 10^{34}$~cm$^{-2}$s$^{-1}$ and an average number of interactions in the same or neighbouring bunch crossings of  $\langle \mu \rangle = 33.7$. With requirements on the stability of the beams, the operational status of all ATLAS detector components, and the quality of the recorded data, the total integrated luminosity of the dataset is 139~\ifb\ with an uncertainty of 1.7\%. It is derived from the calibration of the luminosity scale using $x$--$y$ beam-separation scans, following a methodology similar to that detailed in Ref.~\cite{ATLAS-CONF-2019-021}, and using the LUCID-2 detector~\cite{LUCID2} for the baseline luminosity measurements.
 
To evaluate the effects of the detector efficiency and acceptance on the signal and background, and to estimate SM backgrounds, simulated event samples were produced using Monte Carlo (MC) generators. Inelastic collisions were simulated using \PYTHIA 8.186~\cite{Pythia8} with a set of tuned parameters called the A2 tune~\cite{ATL-PHYS-PUB-2014-021} and the MSTW2008LO parton distribution function (PDF)~\cite{MSTW} set and overlaid on the signal and background MC samples. These simulated events are reweighted to accurately reproduce the average number of proton-proton interactions  in the same or neighbouring bunch crossings (referred to as pile-up).
 
Samples of simulated events for DM production in simplified models were generated  for the case of an $s$-channel mediator with axial-vector interactions.
\MGMCatNLO v2.6.2~\cite{Alwall:2014hca} was used in conjunction with \PYTHIA 8.235 with the A14 tune~\cite{ATL-PHYS-PUB-2014-021}  for modelling of parton showering, hadronisation, and the underlying event.
The PDF set used is NNPDF3.0NLO~\cite{Ball:2014uwa}. A photon with a transverse energy above 130~\GeV\ was required at the matrix-element level in \MGMCatNLO. The DMsimp~\cite{Backovic:2015soa} implementation of the model at next-to-leading order (NLO) in QCD was used. The $g_q$ coupling was set to be universal in quark flavour and equal to 0.25, $g_{\chi}$ was set to 1.0, and
$\Gamma _{m_{\textrm{med}}}$ was computed as the minimum width allowed given the couplings and masses. Different choices of the couplings and a model with a vector mediator were also considered~\cite{DMwg}.  A grid of signal samples was generated spanning the $m_{\chi}$--$m_{\textrm{med}}$ plane for $m_{\chi}$ values from 10 to 500~\GeV\ and $m_{\textrm{med}}$ values from 10 to 1700~\GeV.

Samples for ALP production in association with a photon were generated at NLO with \MGMCatNLO v2.6.2 and interfaced to \PYTHIA 8.240 with the A14 tune for modelling of parton showering, hadronisation, and the underlying event. The PDF set used for the generation is NNPDF23LO, and the renormalisation and factorisation scales were set to half of the transverse mass of the ALP and photon system.
Effective scales $f_a$ in the range between 1~\TeV\ and 5~\TeV\ are explored.
 
Samples of SM backgrounds were produced using the {\SHERPA} 2.2 MC event generator~\cite{Gleisberg:2008ta, Sherpa2}.
The leading-order (LO) and NLO matrix elements were generated using the Comix~\cite{Gleisberg:2008fv} and OpenLoops~\cite{Cascioli:2011va} matrix-element generators, and the merging with the parton shower is done using the ME+PS@NLO prescription~\cite{Hoeche:2009rj}.  The NNPDF3.0NNLO~\cite{Ball:2014uwa} PDF set was used in conjunction with a dedicated parton shower tuning developed by the  {\SHERPA} authors~\cite{Schumann:2007mg}.  For $W\gamma$ and $Z\gamma$ backgrounds, events containing a charged lepton ($e$, $\mu$ or $\tau$) and a neutrino or a pair of charged leptons together with a photon and associated jets were simulated by  {\SHERPA} 2.2.2 and the matrix elements were calculated for up to one parton at NLO and up to three partons at LO.
For $\gamma^*/Z$ decays into charged leptons, a requirement on the dilepton invariant mass of $m_{\ell\ell} > 10$~\GeV\  was applied at generator level. Events containing a photon with associated jets were also simulated using {\SHERPA} 2.2.2,  and the matrix elements were calculated for up to two partons at NLO and up to four partons at LO.  Events containing $W$ or $Z$ bosons with associated jets were simulated using {\SHERPA} 2.2.1 and the matrix elements were calculated for up to two partons at NLO and up to four partons at LO.
The $W/Z$+jets events are normalised to next-to-next-to-leading order (NNLO) inclusive cross-section predictions~\cite{ATL-PHYS-PUB-2016-003}.

Table~\ref{tab:MC}  summarises the details of the generation of  events for the signal samples and SM background processes considered in the analysis.
 
\begin{table}[h!]
\begin{center}
\caption{Details of the generation of events for the signal samples and SM backgrounds considered in the analysis.}
\setlength{\tabcolsep}{0.0pc}
{\small
\begin{tabular*}{\textwidth}{@{\extracolsep{\fill}}lccc}
\noalign{\smallskip}\hline\hline\noalign{\smallskip}
 
Process & Generators  & PDF sets & Order \\
\noalign{\smallskip}\hline\noalign{\smallskip}
DMsimp model 	& \textsc{MG5}\_a\textsc{MC@NLO} v2.6.2 + \PYTHIA  8.235 &     NNPDF3.0NLO & NLO \\
ALP model         & \textsc{MG5}\_a\textsc{MC@NLO} v2.6.2 +   \PYTHIA  8.240 &  NNPDF23LO  & NLO    \\
$W/Z+\gamma$   & {\SHERPA} 2.2.2  &  NNPDF3.0NNLO &     0,1j@NLO + 2,3j@LO                   \\
$\gamma$+jets  & {\SHERPA} 2.2.2  & NNPDF3.0NNLO&         1,2j@NLO + 3,4j@LO                         \\
$W/Z$+jets       & {\SHERPA} 2.2.1  &  NNPDF3.0NNLO &        0,1,2j@NLO + 3,4j@LO                           \\
\noalign{\smallskip}\hline\hline\noalign{\smallskip}
\end{tabular*}
\label{tab:MC}
}
\end{center}
\end{table}

All background samples were simulated with a full ATLAS detector simulation~\cite{:2010wqa} based on \textsc{Geant4}~\cite{Agostinelli:2002hh}, while the signal samples  were processed through a fast simulation of the ATLAS detector using a parameterisation of the calorimeter response and \textsc{Geant4} for the ID and MS.
The simulated events are reconstructed and analysed with the same analysis chain as for the data, using the same trigger and event selection criteria discussed in Section~\ref{sec:EventReco}.

 
\section{Event reconstruction and selection}
\label{sec:EventReco}

All the events in the analysis must satisfy beam, detector and data-quality criteria. They are selected by a trigger requiring at least one photon candidate with transverse energy $\phPT$ above a threshold of $140~\GeV$ and passing `loose' identification requirements~\cite{TRIG}.  Events are required to have at least one candidate for primary vertex, defined as the vertex with the highest sum of the squared transverse momenta of its associated tracks and reconstructed from at least two good-quality tracks ~\cite{ATL-PHYS-PUB-2015-026} with $\pt > 0.5~\GeV$.  Events are removed if they contain a poor-quality photon or jet arising from instrumental problems or non-collision background ~\cite{ATLAS-CONF-2015-029}.
 
Depending on the quality and kinematic requirements imposed, physics objects are labelled either as candidate or selected, where the latter is a subset of the former, with tighter selection criteria applied. Candidate physics objects are used when classifying overlapping selected objects and for vetoing events.
 
Photons are reconstructed from clusters of energy deposits in the electromagnetic calorimeter, together with information about charged-particle tracks reconstructed in the ID~\cite{egamma}.
Photon candidates are classified either as converted (the photon cluster is matched to a reconstructed conversion vertex) or as unconverted (matched to neither a conversion vertex nor an electron track). Both the converted and unconverted photon candidates are used in the analysis.
The calibration of the photon energy in the calorimeter accounts for upstream energy loss as well as lateral and longitudinal leakages. The energy of the cluster of calorimeter cells associated with the photon candidate is corrected using a combination of simulation-based and data-driven calibration factors determined from $Z \rightarrow e^+ e^-$ events.
Photon candidates are also required to fulfill `loose' or `tight' identification criteria based on observables that reflect the shape of the electromagnetic showers in the calorimeter, especially in the finely segmented first layer.
All the reconstructed photons of `loose' quality with $\phPT > 10~\GeV$ and $|\eta| < 2.37$ are considered as photon candidates. Selected photons must additionally satisfy the `tight' identification criteria, have $|\eta| < 1.37$ or $1.52 < |\eta| < 2.37$, and be isolated to avoid contamination coming from $\pi^0$ or other neutral hadrons decaying into an almost-collinear photon pair; the isolation condition is imposed by requiring the transverse energy in the calorimeters in a cone of size $\Delta R = \sqrt{(\Delta \eta)^2 + (\Delta \phi)^2} = 0.4$ around the cluster barycentre, excluding the transverse energy associated with the photon cluster, to be less than $2.45~\GeV + 0.022\phPT$. This transverse energy in the cone is corrected for photon energy leakage from the central core and for the effects of pile-up and the underlying event.
In addition, the scalar sum of the $\pt$ of non-conversion tracks in a cone of size $\Delta R = 0.2$ around the cluster barycentre is required to be less than $0.05\phPT$.
 
Electron candidates are reconstructed from energy clusters in the electromagnetic calorimeter matched to charged-particle tracks in the ID. A procedure similar to that for photons is applied for their identification and calibration~\cite{egamma}. They must fulfil the `medium' identification requirement and are required to have $\pt > 7~\GeV$ and $|\eta| < 2.47$. Selected electrons must be isolated according to `loose' criteria.
 
Muon candidates are reconstructed by searching for track segments in different layers of the MS. These segments are combined and matched with tracks found in the ID. The candidates are re-fitted using the complete track information from both detector systems. Muon candidates must pass the `medium' identification requirement~\cite{PERF-2015-10} and are
required to have $\pt > 6~\GeV$ and $\abseta < 2.7$. Selected muons are also required to pass the `loose' isolation requirement.
 
To achieve additional rejection of background electrons and muons from non-prompt sources, electron and muon tracks must be matched to the primary vertex with a longitudinal impact parameter
$|z_0\sin\theta | < 0.5$~mm and a transverse impact parameter significance, defined as the transverse impact parameter $d_0$ divided by its estimated uncertainty $\sigma_{d_0}$, satisfying $|d_0|/\sigma_{d_0} <$~5.0 for electrons and $|d_0|/\sigma_{d_0} <$~3.0 for muons.
 
Jets are reconstructed from topological clusters of energy in the calorimeter~\cite{topoclusters} using the anti-$k_{t}$ jet clustering algorithm~\cite{Cacciari:2008gp,Fastjet} with a radius parameter $R = 0.4$. The reconstructed jets are then calibrated by the application of a jet energy scale derived from 13~\TeV\ data and simulation~\cite{JES_13TeV}. Only candidate jets with $\pt > 20~\GeV$ and $|\eta| < 4.5$ are considered.
To reduce the effects of pile-up, for jets with $|\eta| < 2.5$ and $\pt < 120~\GeV$ a significant fraction of the tracks associated with each jet must have an origin compatible with the primary vertex, as defined by the jet vertex tagger~\cite{ATLAS-CONF-2014-018}. Selected jets are required to have $\pt > 30~\GeV$ and $|\eta| < 4.5$.
 
Hadronically decaying $\tau$-lepton candidates are reconstructed by combining information from the calorimeters and the ID. The $\tau$-lepton reconstruction algorithm is seeded by jets reconstructed as described above and the reconstructed energy is corrected to the $\tau$-lepton energy scale~\cite{ATLAS-CONF-2017-029}. Hadronically decaying $\tau$-lepton candidates are required to have one or three associated charged-particle tracks (prongs) and  $\pt > 20~\GeV$. To improve the discrimination between hadronically decaying $\tau$-leptons and jets, a multivariate algorithm is used; selected $\tau$-leptons are required to fulfil the  `loose' identification criteria.
 
Possible double counting of reconstructed candidate physics objects is resolved in the following order. If any electron shares its inner detector track with a selected muon, the electron is removed and the muon is kept, in order to remove electron candidates originating from muon bremsstrahlung followed by photon conversion. If a photon and an electron or a muon are closer than $\Delta R = 0.4$ the photon is removed. If an electron lies within $\Delta R = 0.2$ of a jet, the jet is removed, while if an electron lies within $0.2 < \Delta R < 0.4$ of a jet, the electron is removed. Muons lying within $\Delta R = 0.4$ of jets are removed, except if the number of tracks with $\pt > 0.5~\GeV$ associated with the jet is less than three. In the latter case, the muon is kept and the jet is discarded. If a jet lies within $\Delta R = 0.4$ of a photon, the jet is removed. Hadronically decaying $\tau$-leptons close to electrons or muons ($\Delta R < 0.2$) are removed. Any remaining jet within $\Delta R = 0.2$ of a hadronically decaying $\tau$-lepton is removed.
 
The missing transverse momentum \metv\ is measured as the negative vectorial sum of the transverse momenta of all candidate electrons, photons, hadronically decaying $\tau$-leptons, jets, and muons, plus an additional `soft term'~\cite{MET2}.  The `soft term' is constructed from high-quality charged-particle tracks associated with the primary vertex but not with such physics objects. This allows the \metv\ calculation to adopt the best calibration for all the identified particles, while maintaining pile-up independence in the `soft term'. 
Possible double counting of contributions from reconstructed charged-particle tracks, energy deposits in the calorimeter, and reconstructed muons is avoided by applying a signal ambiguity resolution procedure which rejects already used signals when combining the various \metv\ contributions.
A powerful quantity used to discriminate between events with \metv\ arising from poorly reconstructed physics objects and events with \metv\ originating from weakly interacting particles is the $\met$ significance, which is calculated as |\,\metv\,|$/\left[\sigma_\text{L}^2\left(1-\rho_\text{LT}^2 \right) \right]^{1/2}$,
where $\sigma_\text{L}$ is the total standard deviation in the direction longitudinal to the \metv\ corresponding to the summation of the covariance matrices from resolution effects of physics objects and  `soft term' entering the \metv\ calculation, $\rho_\text{LT}$ is the correlation factor of the longitudinal (L) and transverse (T) measurements~\cite{ATLAS-CONF-2018-038}.

The signal region (SR) is defined by requiring events to have a selected leading photon that satisfies the criteria defined in Section~\ref{sec:EventReco} and has $\phPT > 150~\GeV$.
 
The `photon pointing', $|\Delta z_\gamma|$, defined as the separation, measured along the beam line, between the extrapolated origin of the photon and the position of the event's identified primary vertex is required to be smaller than 250~mm.
This criterion suppresses the non-collision background  in the data-driven method used to estimate the contribution from events in which jets are misidentified as photons, as described in Section~\ref{sec:results}.
 
To ensure that the leading photon and \metv\ do not overlap in the transverse plane, $\Delta\phi(\gamma,~$\metv$) > 0.4$ is required. In order to reduce the number of background events characterised by \metv\ arising from poorly reconstructed physics objects, events are required to have $\met > 200$~\GeV\ and $\met$ significance > 8.5.
To suppress multi-jet background, events with more than one selected jet or with a jet with $\Delta\phi(\jet,~$\metv$)<0.4$ are rejected.
Events are required to have no candidate electrons, muons or hadronically decaying $\tau$-leptons passing the requirements described in Section~\ref{sec:EventReco}. This lepton veto mainly rejects $W/Z$ events.
 
To improve the sensitivity of the analysis, seven SRs are defined corresponding to different $\met$ ranges: four inclusive (SRI1--SRI4) and three exclusive (SRE1--SRE3), their use will be explained in Section~\ref{sec:BGEstimation}. Table~\ref{tab:SR_criteria} summarises the SR definitions and the selection criteria described above.

\begin{table}[h!]
\caption{Selection criteria for the SRs.}
\begin {center}
\begin{tabular}{llllllll}
 
\toprule
\toprule
\multicolumn{1}{l}{Event cleaning} & \multicolumn{7}{c}{Detector quality conditions and primary vertex}  \\
&                        &                                                                        &                                                      &                                              &\\[-0.8em]
\multicolumn{1}{l}{Leading $\gamma$} &\multicolumn{7}{c}{$\phPT > 150~\GeV$, $|\eta|<1.37$ or $1.52<|\eta|<2.37$,}  \\
\multicolumn{1}{c}{} &\multicolumn{7}{c}{tight, isolated,  $|\Delta z_\gamma|$  < 250 mm,}  \\
\multicolumn{1}{c}{} &\multicolumn{7}{c}{$\Delta\phi(\gamma,~$\metv$) > 0.4$} \\
&                        &                                                                        &                                                      &                                              &\\[-0.8em]
\multicolumn{1}{l}{$\met$} significance &\multicolumn{7}{c}{$ > 8.5$} \\
&                          &                                                                        &                                                      &                                             &\\[-0.8em]
\multicolumn{1}{l}{Jets} &\multicolumn{7}{c}{0 or 1  with  \pt\ $>30~\GeV$, $|\eta| < 4.5$ and $\Delta\phi(\jet,~$\metv$)>0.4$} \\
&                          &                                                                        &                                                      &                                             &\\[-0.8em]
\multicolumn{1}{l}{Leptons} &\multicolumn{7}{c}{veto on $e$, $\mu$ and $\tau$} \\
\midrule
 
& \multicolumn{1}{c}{ SRI1}&  \multicolumn{1}{c}{SRI2}& \multicolumn{1}{c}{SRI3}& \multicolumn{1}{c}{SRI4}& \multicolumn{1}{c}{SRE1}  & \multicolumn{1}{c}{SRE2} & \multicolumn{1}{c}{SRE3}  \\
 
\multicolumn{1}{l}{$\MET$ [\GeV] }  & \multicolumn{1}{c}{$ > 200$} & \multicolumn{1}{c}{ $ > 250$ } &  \multicolumn{1}{c}{$>300$} &  \multicolumn{1}{c}{$>375$} &  \multicolumn{1}{c}{200--250} &  \multicolumn{1}{c}{250--300} & \multicolumn{1}{c}{300--375} \\
&                          &                                                                        &                                                      &                                             &         &  &  \\[-0.8em]
 
\bottomrule
\bottomrule
\end{tabular}
\end {center}
\label{tab:SR_criteria}
\end{table}

 
\section{Strategy for background estimation \label{sec:BGEstimation}}
 
Backgrounds in the various SRs arise from a number of processes that generate real photons and from events in which one or more energetic jets or electrons are misidentified as photons.
The latter are estimated through the use of control samples including jets or electrons, scaled by misidentification rates determined from data.
The background where isolated photons are accompanied by significant  \metv\ is expected to receive contributions from processes with energetic neutrinos providing genuine \metv\ and from events where the \metv\ arises from instrumental sources or poorly reconstructed physics objects.
These contributions are obtained using MC simulations constrained by observed event counts in dedicated control regions (CRs) through the estimation of normalisation factors.
Control regions are built by inverting one or more of the selection criteria used to define the SRs, allowing one of the background processes to become dominant but otherwise kinematically similar to the given SR.
The contribution of these backgrounds is determined separately for each SR, as defined in Table~\ref{tab:SR_criteria}, with a maximum-likelihood fit,
referred to as the `background-only fit'.
This procedure constrains the normalisation of the dominant backgrounds to the observed event yields in the associated CRs, assuming that no signal is present in the CRs and including the background
contribution from the misidentification of electrons and jets as photons estimated with data-driven techniques.
 
The inputs to the fit for each SR include the number of events observed in its associated CRs and the number of events predicted by simulation in the SR and CRs for each background process.  The latter are described by Poisson statistics.
The systematic uncertainties in the expected values (see Section~\ref{sec:results})  are included in the fit as nuisance parameters, modelled by Gaussian distributions with widths corresponding to the sizes of the associated uncertainties.
Using a simultaneous fit technique allows a straightforward combination of multiple CRs and permits a coherent treatment of the correlation of the systematic uncertainties across the different regions.
The product of the various probability distributions forms the likelihood, which the fit maximises by adjusting the background normalisations and the nuisance parameters. 
 
Moreover, a simultaneous fit is performed using the CRs associated to the exclusive SRs  plus the inclusive SR corresponding to the highest  $\met$ range (SRI4) as shown in Table~\ref{tab:SR_criteria}.
This allows normalisation factors for the background contributions in each $\met$ bin to be extracted by exploiting the $\met$  shape information. Since the shape of BSM signal models over multiple $\met$ bins is different from the background prediction, this technique,  known as `simplified shape fit', permits to better discriminate signal from background and it is used to set exclusion limits in the models studied, if no excess is found in the data.

Four control regions are defined in order to estimate the contributions of the dominant $\gammaznunu$ background and secondary $\wgamma$, $\zgamma$ and $\gamma+\jets$ backgrounds, making use of the maximum-likelihood approach described above.
 
The Single-Muon CR  used to extract the normalisation of the $W(\rightarrow \ell \nu) \gamma$ background in the corresponding signal regions is built by selecting events with the same criteria used for each SR, except for the muon veto. Exactly one selected muon must be present in each event.
In addition, since background events characterised by fake \metv\ arising from poorly reconstructed jets are not expected to contribute significantly in the leptonic CRs, the requirement on the $\met$ significance is not applied.
For this control region the \metv\ is defined as described in Section~\ref{sec:EventReco}  but the muon contribution is
not taken into account in the computation to emulate  the $\met$ distribution in the SR.  A similar Single-Electron CR is not needed to constrain the $W(\rightarrow \ell \nu) \gamma$ normalisation because the Single-Muon CR has enough events and is less contaminated by events with fake photons and fake \metv.
 
To constrain the normalisation of both the $\gammaznunu$ and the $Z(\rightarrow \ell \ell) \gamma$ processes in each SR, two control regions are defined similarly, with the corresponding $\met$ criteria but inverting the lepton veto.
In the $Z$-enriched Two-Muon and Two-Electron CRs, exactly two selected muons/electrons are required in the event, with a dilepton invariant mass $m_{\ell\ell}$  greater than 10~\GeV. Similarly to the Single-Muon CR, the \metv\ in the Two-Muon CR and the Two-Electron CR is computed  disregarding the contributions from selected muons or electrons, respectively.
 
The $\gamma+\jets$ background in the SRs consists of events characterised by fake \metv\  originating from  jet energy mismeasurements or wrong jet-to-vertex matching amplified by the high pile-up conditions.
This QCD  background is largely suppressed by the large $\met$ requirement, the jet--\metv\ azimuthal separation and the requirement on the $\met$ significance described in Section~\ref{sec:EventReco}.
The control region used to estimate the normalisation of the residual $\gamma+\jets$ background (Photon--Jet CR) is defined by lowering the $\met$ requirement to the range $85~\GeV < \met < 110~\GeV$ and removing the requirement on the $\met$ significance.
In addition, the requirement $\Delta\phi(\gamma,~$\metv$) < 3.0$ is applied to reduce possible signal contamination, thus providing a region dominated by real photons arising from radiative QCD processes.

The background contribution from events in which jets are misidentified as photons, mainly due to $Z$~+~jet or $W$~+~jet processes, is estimated using a sideband counting method~\cite{Aad:2010sp}.
This method relies on counting photon candidates in four regions of a two-dimensional plane defined by the amount of transverse energy deposited in cell-clusters within a cone of size $\Delta R =  0.4$ around the photon, excluding the photon cluster itself (isolation), and by the quality of the photon identification criteria (tightness).
A photon signal region (region A) is defined by photon candidates that are isolated and satisfy `tight' identification as explained in Section~\ref{sec:EventReco}.
Three background regions are defined in the isolation--tightness plane, consisting of photon candidates which are tight and non-isolated (region B), non-tight and isolated (region C) or non-tight and non-isolated (region D).
A non-isolated photon candidate is defined by inverting the requirement in the amount of isolation transverse energy.  
A photon candidate is classified as non-tight if it  fails the tight identification but satisfies a modified set of requirements related to four of the selections associated with the shower-shape variables computed from the energy deposits in the first layer of the EM calorimeter.
Complete independence between the photon identification and isolation would imply that  the numbers of  photon candidates in the four regions  (A, B, C, D) satisfy the condition $N^\mathrm{A} / N^\mathrm{B} = N^\mathrm{C}/ N^\mathrm{D}$.
Although this condition is almost fully satisfied, to estimate the number of background candidates in the region A there is a residual correlation that has to be taken into account.
Besides, a correction to the method is added in order to consider the effect of contamination by real photon events in the three background regions (B, C, D).
MC simulations are used to estimate both the correlation factor and the signal leakage coefficients.
This method is then used to evaluate the contribution of jets misidentified as photons in all analysis regions: the SRs and their associated four CRs used for the dominant background.
 
Electrons or positrons can be misidentified as photons and represent an additional source of background. This contribution is estimated by using a control sample with an electron--$\met$ final state and scaling the event yield by the probability for such an electron to be misreconstructed as a tight photon as determined from a comparison of the rate of $Z$ boson reconstruction in the $e\gamma$ and $ee$ final states.
The full Run-2 dataset is used to select $Z \rightarrow ee$ events where the two electrons (actually, one of the two electrons is a positron,  but they are referred to as electrons in what follows) in the final state are reconstructed either as an $ee$ pair   or as an $e\gamma$ pair. The invariant mass $m_{ee}$ or $m_{e\gamma}$ is required to be consistent with the $Z$ boson mass to within 10~\GeV.
The yields of $Z$ events are then obtained from a fit to the $m_{ee}$ and $m_{e\gamma}$ distributions, in order to subtract the contamination from misidentified jets in the sample as modelled from the  sidebands.
The electron-to-photon scale factor, measured as a function of $|\eta|$ and $\pt$ varies between 1.5\% and 9\%, with larger factors associated with larger values of $|\eta|$, since the misidentification rate depends on the amount of material in front of the calorimeter.
Background estimates are then also made for  the various signal regions as well as for their associated four control regions by
applying  the electron-to-photon misidentification factor to events selected with the same criteria as used in these regions but requiring an electron instead of a photon.

 
\section{Results }\label{sec:results}
 
The background-only simultaneous fit,  described in Section~\ref{sec:BGEstimation}, is performed to evaluate the SM background expectations in all regions of the analysis.
 
Systematic uncertainties in each of the background components are taken into account in the fit as described in Section~\ref{sec:BGEstimation}.
Uncertainties arising from experimental and theoretical sources are estimated for the $\gammaznunu$, $W(\rightarrow \ell \nu) \gamma$, $Z(\rightarrow \ell \ell) \gamma$ and $\gamma+\jets$ backgrounds.
Experimental uncertainties related to the energy and momentum scale of all the physics objects described in Section~\ref{sec:EventReco}, and to their identification, reconstruction and isolation efficiencies, are taken into account.
The uncertainties of the different physics objects are propagated to the \met calculation, as are the uncertainties in the `soft term' resolution and scale.
The uncertainty in the integrated luminosity and pile-up reweighting, reported in Section~\ref{sec:MCSimulations}, is also considered.
 
Theoretical uncertainties affecting the {\SHERPA} MC event generator predictions include an uncertainty  in the NLO cross section as well as  uncertainties from variations of the 
QCD factorisation and renormalisation scales~\cite{Bothmann:2016nao}, in the strong coupling constant $\alphas$, and from the choice of parton distribution functions. The effects of the latter `PDF+$\alphas$' uncertainties are calculated using the PDF4LHC prescription~\cite{Botje:2011sn}. The theoretical uncertainties affect the dominant $W/Z+\gamma$ backgrounds by less than about 7\%.
 
Both the experimental and theoretical systematic uncertainties are considered fully correlated between the different control and signal regions allowing a partial cancellation in the fitting procedure.

Uncertainties in the estimates of jets and electrons misidentified as photons are evaluated with the in situ techniques explained in Section~\ref{sec:BGEstimation}. The main systematic uncertainty for the sideband counting method is evaluated by varying the criteria for photon tightness and isolation used to define the four regions (A,B,C,D). Uncertainties related to signal leakage were considered in order to account for differences between $W \gamma$ and  $Z \gamma$ samples.  The sources of systematic uncertainties related to the methodology used to estimate the contribution of electrons misidentified as photons are the choice of invariant mass window, the background subtraction and the energy scale of the misidentified photons; the last of these has the dominant effect.

Table~\ref{tab:SRI1_yields} shows the observed number of events and the expected SM background contribution in SRI1, which is the most inclusive SR and covers $\met > 200$~\GeV, and in its associated CRs: three leptonic CRs and the Photon--Jet CR.  For the SM predictions, both the statistical and systematic uncertainties are included.
 
The observed number of events and the expected SM background contribution resulting from the  background-only fit  are reported in Table~\ref{tab:SRI_yields}  for each inclusive SR.
The values of the normalisation factors for  $W/Z+\gamma$  and $\gamma+\jets$ backgrounds ($k_{Z\gamma}$, $k_{W\gamma}$, $k_{\gammajets}$) are also shown.
A summary of the data event yields and the background expectations broken down into their contributing SM sources in the inclusive regions is shown in Figure~\ref{fig:summary_inclusive}, which also presents the significance of the difference between data and the SM background prediction, calculated with the method described in Ref.~\cite{Casadei}.
 
Results from the `simplified shape fit' described in Section~\ref{sec:BGEstimation}, are shown in Table~\ref{tab:SRE_yields} together with the fitted normalisation factors.
The fit is performed simultaneously in the CRs associated with each of the exclusive SRs and with the inclusive  SR with the highest \met range (SRI4).
The Photon--Jet CR is common to all SRs as it corresponds to a lower \met range, while the leptonic CRs span the same \met ranges of the SRs involved.
A summary of the data event yields and the SM expectations  in all the regions used for the `simplified shape fit' to extract the model-dependent limits is shown in Figure~\ref{fig:summary}.
 
The distributions of $\met$  in data and for the expected SM background obtained after performing the `simplified shape fit'  are shown in Figure~\ref{fig:postfit} for the SRs and the leptonic CRs.
The distributions of the expected SM backgrounds include the $W/Z + \gamma$ and $\gamma+\jets$ backgrounds  normalised with the corresponding $k$-factors reported in Table~\ref{tab:SRE_yields} and the data-driven estimates for the backgrounds produced by events in which electrons or jets are misidentified as photons.

\begin{table}[!htbp]
\begin {center}
\small\addtolength{\tabcolsep}{-0.2pt}
\caption{
Observed and expected yields from SM backgrounds in SRI1 corresponding to $\met > 200$~\GeV\ and in its associated four CRs.
The expected event yields from SM processes are obtained from the background-only fit, described in Section~\ref{sec:BGEstimation}.
The uncertainty includes both the statistical and systematic uncertainties. The individual uncertainties can be correlated
and do not necessarily add in quadrature to equal the total background uncertainty.
}
\label{tab:SRI1_yields}
\begin{tabular}{cccccc}
\toprule
\toprule
& SRI1 & 1 Muon CR & 2 Muon CR & 2 Electron CR & Photon--Jet CR \\
\midrule
Observed events& $5293$& $1991$& $473$& $378$& $21991$\\
\midrule
Expected SM events& $5350 \pm 190$& $1991 \pm 45$& $475 \pm 18$& $376 \pm 15$& $21990 \pm 150$\\
\midrule
$Z(\to\nu\nu)\gamma$& $3410 \pm 150$& $~~~~1.721 \pm 0.094$&~~--  &~~--  & $~~372 \pm 42$\\
$W(\to\ell\nu)\gamma$& $680 \pm 33$& $1589 \pm 68$& $~~0.40 \pm 0.11$& $~~0.81 \pm 0.18$& $~~530 \pm 37$\\
$Z(\to\ell\ell)\gamma$& $48.4 \pm 2.9$& $131.5 \pm 8.1$& $457 \pm 19$& $361 \pm 16$& $~~35.9 \pm 4.8$\\
$\gamma$ + jets& $103 \pm 41$& $~~12.9 \pm 7.0$& ~~--  & --& $19610 \pm 290$\\
Fake photons from $e$& $860 \pm 80$& $~~63.9 \pm 6.0$& $~~1.91 \pm 0.31$& $~~0.54 \pm 0.22$& $~~694 \pm 65$\\
Fake photons from jets& $249 \pm 54$& $~~192 \pm 49$& $15.3 \pm 8.6$& $13.7 \pm 8.5$& $~~~~750 \pm 230$\\
 
\bottomrule
\bottomrule
\end{tabular}
\end {center}
\end{table}

\begin{table}[!htbp]
\begin {center}
\caption{
Observed and expected yields from SM backgrounds in all inclusive SRs.
The expected event yields from SM processes are obtained from the background-only fit, described in Section~\ref{sec:BGEstimation}, in each inclusive SR.
The normalisation factors obtained from the fit are also shown.
The uncertainty includes both the statistical and systematic uncertainties. The individual uncertainties can be correlated
and do not necessarily add in quadrature to equal the total background uncertainty.
}
\label{tab:SRI_yields}
\begin{tabular}{ccccc}
 
\toprule
\toprule
& SRI1 & SRI2 & SRI3 & SRI4 \\
\midrule
Observed events & $5293$  & $2270$  & $1106$  & $427$  \\
\midrule
Expected SM events& $5350 \pm 190$& $2320 \pm 110$& $1134 \pm 71~~$& $448 \pm 42$\\
\midrule
$Z(\to\nu\nu)\gamma$& $3410 \pm 150$& $1540 \pm 95~~$& $779 \pm 65$& $306 \pm 40$\\
$W(\to\ell\nu)\gamma$& $680 \pm 33$& $285 \pm 22$& $128 \pm 12$& $56.8 \pm 7.1$\\
$Z(\to\ell\ell)\gamma$& $48.4 \pm 2.9$& $16.4 \pm 1.2$& $~~7.01 \pm 0.65$& $~~2.69 \pm 0.37$\\
$\gamma$ + jets& $103 \pm 41$& $17.0 \pm 7.0$& $~~5.5 \pm 2.2$& $~~2.9 \pm 1.2$\\
Fake photons from $e$& $860 \pm 80$& $349 \pm 32$& $161 \pm 15$& $59.7 \pm 5.6$\\
Fake photons from jets& $249 \pm 54$& $114 \pm 40$& $~~54 \pm 20$& $~~20 \pm 11$\\
\midrule
$k_{Z\gamma}$& $~~0.93 \pm 0.07$& $ ~~0.88 \pm 0.07$& $~~0.88 \pm 0.09$& $~~0.86 \pm 0.12$\\
$k_{W\gamma}$& $~~0.81 \pm 0.08$& $~~0.80 \pm 0.09$& $~~0.78 \pm 0.10$& $~~0.85 \pm 0.13$\\
$k_{\gamma + \mathrm{jets}}$& $~~0.82 \pm 0.21$& $~~0.82 \pm 0.20$& $ ~~0.81 \pm 0.19$& $~~0.82 \pm 0.21$\\

\bottomrule
\bottomrule
\end{tabular}
\end {center}
\end{table}

\begin{table}[!htbp]
\begin {center}
\caption{
Observed and expected yields from SM backgrounds obtained from the  `simplified shape fit' described in Section~\ref{sec:BGEstimation}.
The normalisation factors obtained from the fit are also shown.
The uncertainty includes both the statistical and systematic uncertainties. The individual uncertainties can be correlated
and do not necessarily add in quadrature to equal the total background uncertainty.
}
\label{tab:SRE_yields}
\begin{tabular}{ccccc}
\toprule
\toprule
& SRE1 & SRE2 & SRE3 & SRI4 \\
\midrule
Observed events& $3023$& $1164$& $679$& $427$\\
\midrule
Expected SM events& $3070 \pm 130$& $1182 \pm 75~~$& $680 \pm 53$& $448 \pm 42$\\
\midrule
$Z(\to\nu\nu)\gamma$& $1910 \pm 110$& $758 \pm 65$& $468 \pm 49$& $306 \pm 40$\\
$W(\to\ell\nu)\gamma$& $394 \pm 22$& $159 \pm 15$& $71.0 \pm 8.2$& $56.7 \pm 7.1$\\
$Z(\to\ell\ell)\gamma$& $33.2 \pm 2.4$& $~~9.32 \pm 0.89$&~ $4.26 \pm 0.48$& $~~2.69 \pm 0.37$\\
$\gamma$ + jets& $~~87 \pm 35$& $11.9 \pm 4.8$& $~~2.7 \pm 1.1$&  ~~$3.0 \pm 1.2$\\
Fake photons from $e$& $511 \pm 48$& $188 \pm 18$& $100.9 \pm 9.5$~~~& $59.7 \pm 5.6$\\
Fake photons from jets& $136 \pm 28$& $~~56 \pm 29$&~~ $33 \pm 16$&~~ $20 \pm 11$\\
\midrule
$k_{Z\gamma}$& $~~0.99 \pm 0.08$&~~ $0.89 \pm 0.09$&~~ $0.90 \pm 0.11$& ~~$0.86 \pm 0.12$\\
$k_{W\gamma}$& $~~0.81 \pm 0.09$& ~~~$0.84 \pm 0.11$& ~~~$0.74 \pm 0.11$& ~~$0.85 \pm 0.13$\\
$k_{\gamma+\mathrm{jets}}$ & \multicolumn{4}{c}{$0.82 \pm 0.21$} \\

\bottomrule
\bottomrule
\end{tabular}
\end {center}
\end{table}

\begin{figure}[!htbp]\centering
\includegraphics[width=0.9\textwidth]{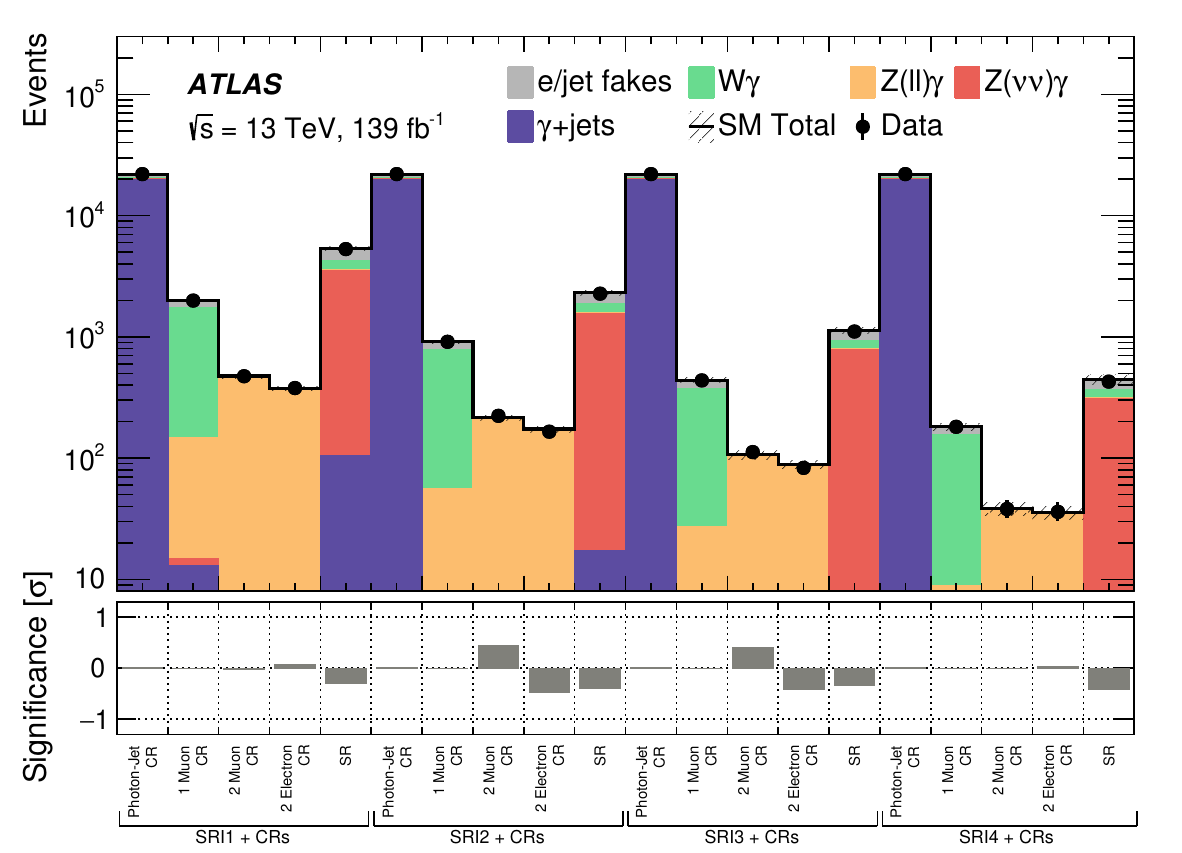}
\caption{
Data event yields and the SM predictions from separate background-only fits in each inclusive SR and  its associated CRs. The uncertainties in the expected numbers of events are the combined statistical and systematic uncertainties.
The lower panel shows the significance of the difference between data and background prediction.}
\label{fig:summary_inclusive}
\end{figure}

\begin{figure}[!htbp]\centering
\includegraphics[width=0.9\textwidth]{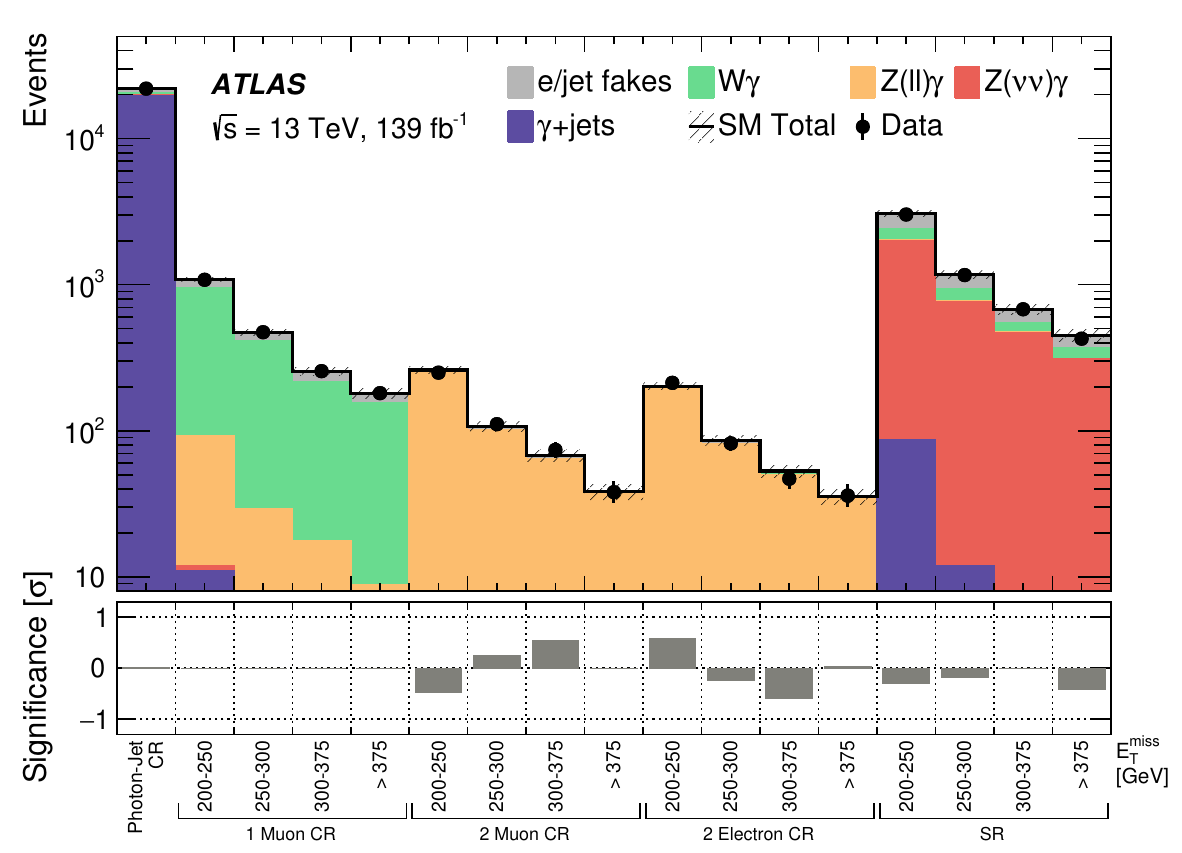}
\caption{
Data event yields and the SM predictions from the `simplified shape fit' in all  exclusive SRs plus the inclusive SR with the highest \met range (SRI4) and their associated CRs. The uncertainties in the expected numbers of events are the combined statistical and systematic uncertainties.
The lower panel shows the significance of the difference between data and background prediction.
}
\label{fig:summary}
\end{figure}

\begin{figure}[!htbp]\centering
\includegraphics[width=6cm,height=6cm]{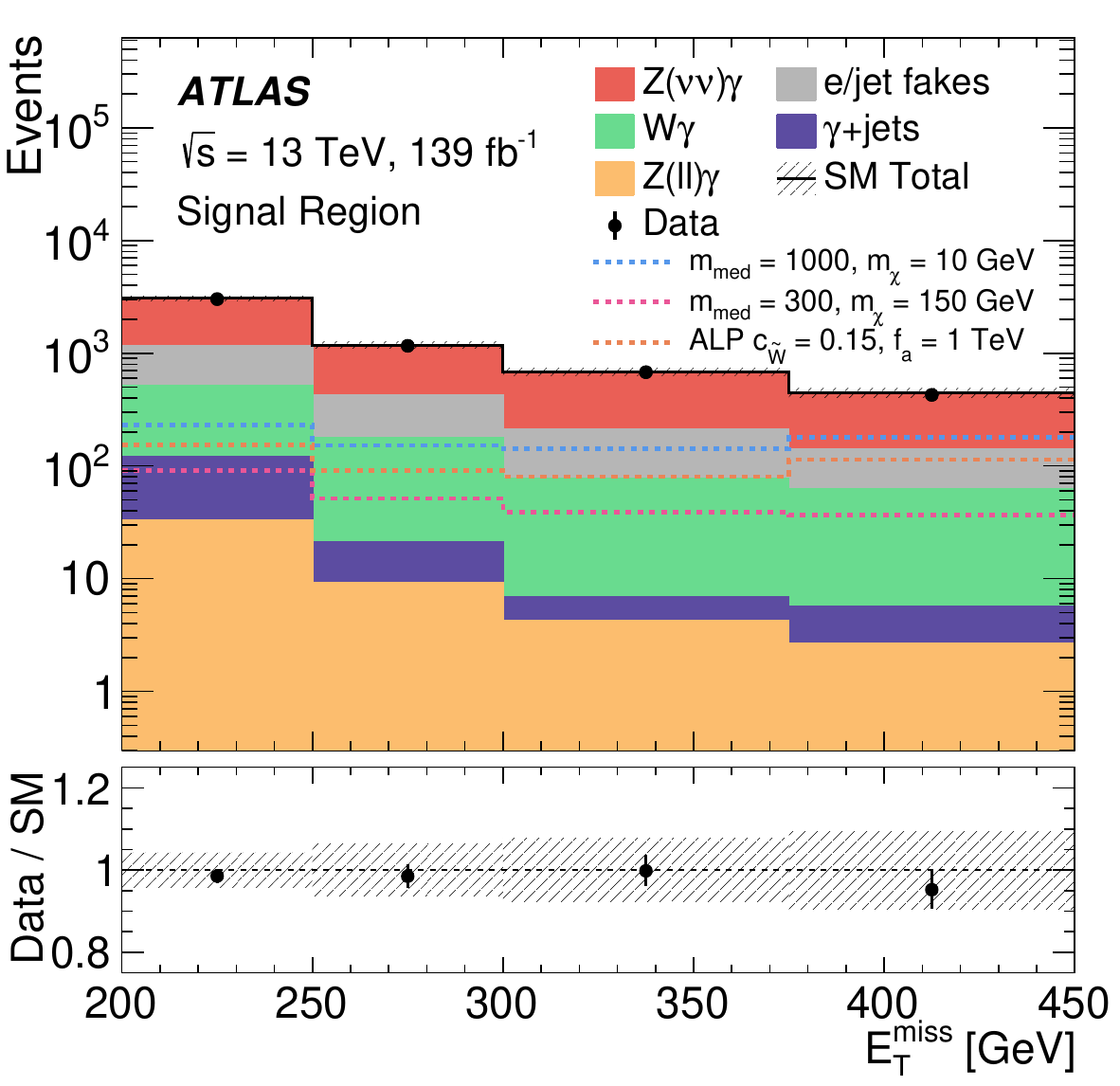}
\includegraphics[width=6cm,height=6cm]{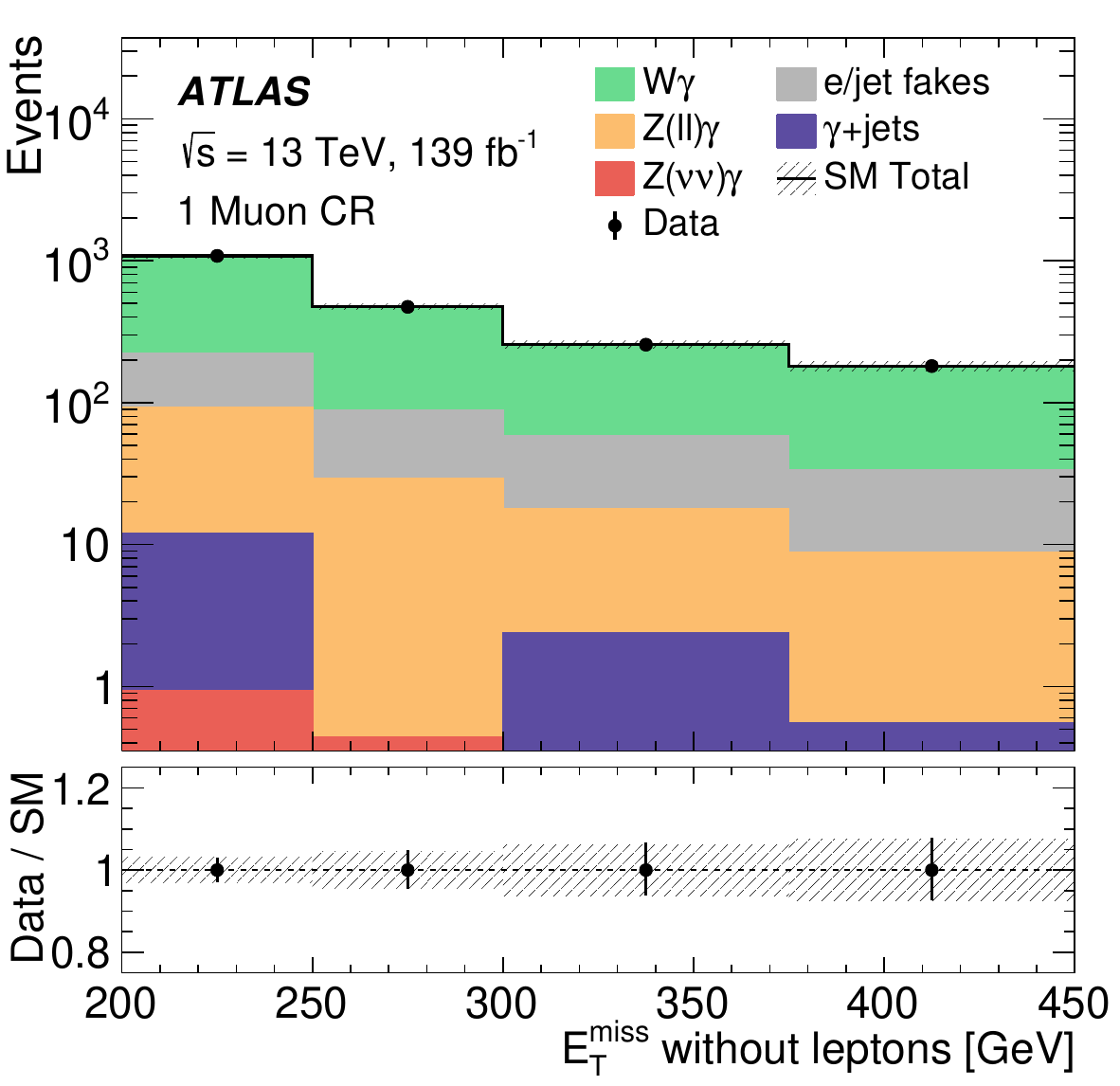}\\
\includegraphics[width=6cm,height=6cm]{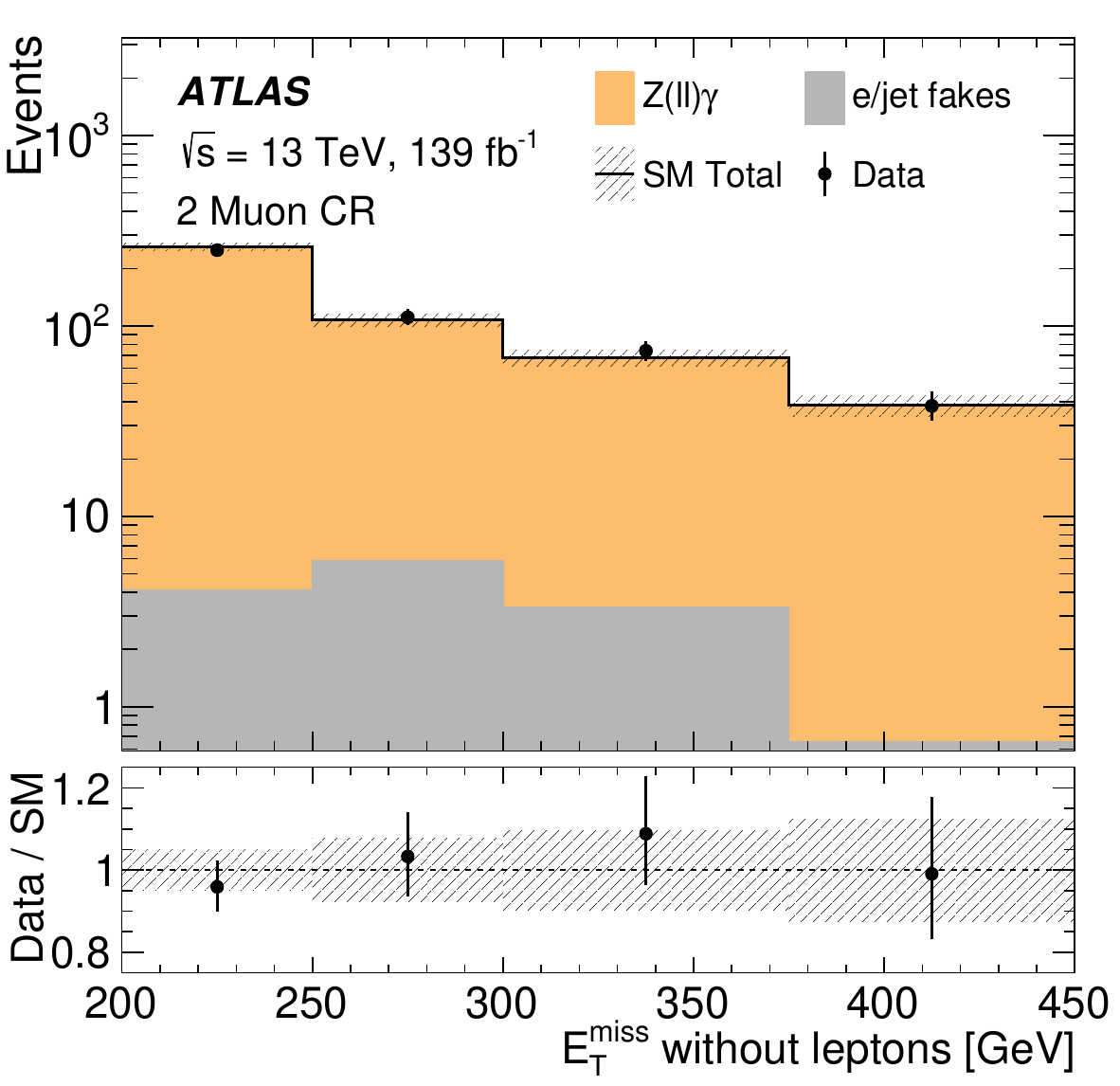}
\includegraphics[width=6cm,height=6cm]{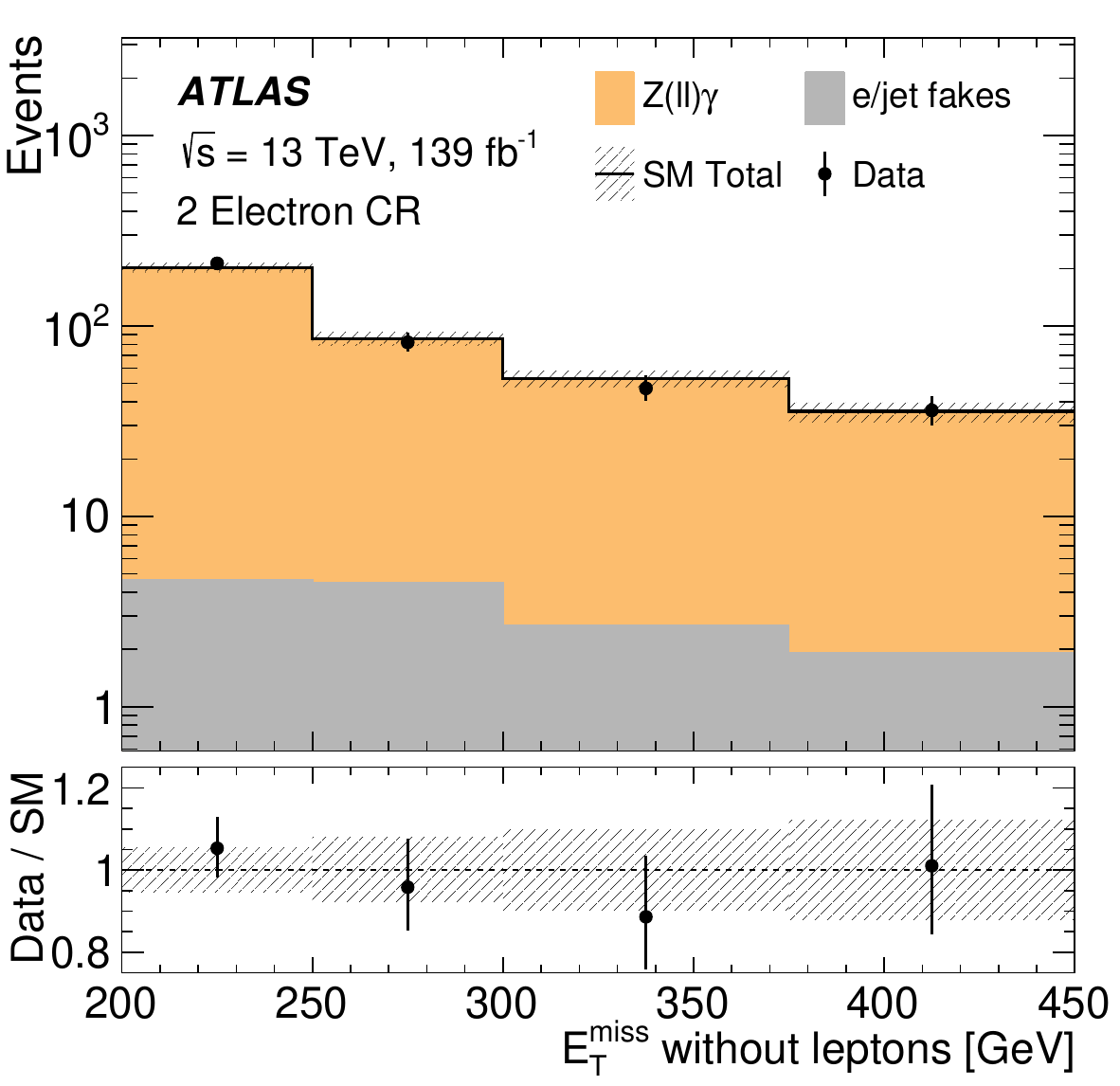}
 
\caption{
Distribution of $\met$ in data and for the expected SM background in the SRs and CRs after performing the `simplified shape fit':
SRs (top left), Single-Muon CR (top right), Two-Muon CR (bottom left) and Two-Electron CR (bottom right). Overflows are included in the fourth bin of each distribution.
The $\met$ calculation in these CRs does not include  the muon or electron contribution.
The error bars are statistical, and the dashed band includes statistical and systematic uncertainties determined by the fit.
The expectations for the simplified model for two different values of $m_{\chi}$ and $m_{\textrm{med}}$, and with $g_{q}=0.25$ and $g_{\chi}=1.0$ and for the ALP model are also shown.
The lower panel shows the ratio of data to expected background event yields.
}
\label{fig:postfit}
\end{figure}

Table~\ref{tab:syst} shows the total relative uncertainty, including systematic and statistical contributions, in the expected SM background yield after the background-only fit for inclusive SRs  and after the `simplified shape fit' for exclusive SRs. This total uncertainty ranges from 3.5\% to 9.5\% depending on the SR.
The purely statistical uncertainty is dominant, varying from 2.4\% to 8.5\% and driven by the statistical precision from the Two-Muon and Two-Electron CRs adopted to constrain the normalisation of the leading $Z(\rightarrow \nu\nu)\gamma$ background.
 
The relative impact of each source of systematic uncertainty on the total SM background estimates is summarised in Table~\ref{tab:syst}.
A large experimental uncertainty is related to jets misidentified as photons and varies from 1.4\% to 4.1\%. The impact of the uncertainty in the jet energy scale and resolution varies from 1.6\% to 2.7\%. The uncertainty related to electrons misidentified as photons varies between about 2\% and 2.3\%.
Other experimental systematic uncertainties related to electrons, photons, muons and the $\met$ `soft term'  have a relative impact below 1.5\% in all SRs.
Theoretical systematic uncertainties in the $W/Z + \gamma$ and $\gamma+\jets$ MC estimates have an impact of 0.5\% in all SRs.
 
In all SRs of the analysis, the observations and background predictions are found to be compatible within the uncertainties.

\begin{table}[!hbtp]
\centering
\small
\caption{
Summary of the uncertainties (\%) in the background estimate for inclusive SRs after the background-only fit and for exclusive SRs after the `simplified shape fit'.
The individual uncertainties can be correlated and do not necessarily add in quadrature to equal the total background uncertainty.
}
\label{tab:syst}
\begin{tabular}{llllllll}
\hline
\hline
& SRI1& SRI2 & SRI3 & SRI4 & SRE1 &SRE2 &SRE3  \\
& [\%]  & [\%] & [\%] & [\%] & [\%] & [\%] & [\%]  \\
 
\hline
Total (statistical+systematic) uncertainty                     & 3.5 & 4.8 & 6.2 & 9.5 & 4.3 & 6.3 & 7.8 \\
Statistical uncertainty                                        & 2.4 & 3.6 & 5.3 & 8.5 & 3.3 & 5.0 & 6.7 \\
\hline
Fake photons from jets (Section~\ref{sec:BGEstimation})        & 1.4 & 2.5 & 2.8 & 4.1 & 1.4 & 3.6 & 3.7 \\
Jet energy scale/resol ~\cite{JES_13TeV}                           & 1.6 & 2.2 & 2.5 & 2.7 & 2.2 & 2.2 & 2.3 \\
Fake photons from electrons  (Section~\ref{sec:BGEstimation})  & 2.1 & 2.0 & 2.0 & 2.0 & 2.3 & 2.3 & 2.1 \\
Electrons	reco/id/isolation eff.	~\cite{egamma}	 	   & 1.0 & 1.2 & 1.3 & 1.4 & 1.0 & 1.0 & 1.2 \\
Electron/photon energy scale/resol       ~\cite{egamma} & 0.8 & 0.6 & 0.7 & 0.9 & 0.9 & 0.9 & 0.6 \\
Muon reco/id/isolation eff.	~\cite{PERF-2015-10}               & 0.7 & 0.8 & 0.9 & 1.0 & 0.6 & 0.7 & 0.9 \\
\met    soft term scale/resolution	~\cite{MET2}               & 0.1 & 0.4 & 0.7 & 0.9 & 0.5 & 0.2 & 0.5 \\
Theoretical $W/Z + \gamma$, $\gamma+\jets$                     & 0.5 & 0.5 & 0.5 & 0.5 & 0.5 & 0.5 & 0.5 \\
$\langle \mu \rangle$ reweighting in MC simulation              & 0.3 & 0.2 & 0.2 & 0.2 & 0.3 & 0.2 & 0.2 \\

\hline
\hline
\end{tabular}
\end{table}

 
\section{Interpretations}
For each SR, exclusion upper limits at the 95\% confidence level (CL) are set on the number of events from any scenario of physics beyond the SM that would produce an excess in events with a $\gam + \met$ final state as presented in this paper.  These limits are based on the profile-likelihood-ratio test statistic~\cite{HistFitter} and CL$_\textrm{s}$ prescriptions~\cite{0954-3899-28-10-313}, evaluated using the asymptotic approximation~\cite{Cowan:2010js}.
A  simultaneous fit is performed including both the SR and its associated CRs to obtain the model-independent upper limits on the observed number of such events.
Normalising the upper limits on the number of signal events by the integrated luminosity of the data sample provides upper limits on the visible BSM cross section  $\sigma\times A\times\epsilon$.
Here $\sigma$ is the production cross section for the BSM signal, and $A\times\epsilon$ is the product of the acceptance ($A$), defined to be the fraction of events whose underlying objects pass all kinematic selections at the particle level, and the efficiency ($\epsilon$), defined to be the fraction of those events that would be observed after reconstruction in the detector.
The expected (in absence of new physics) and observed 95\% CL limits on $\sigma\times A\times\epsilon=N^\text{lim}/\int L\textrm{d}t$ are shown in Table \ref{tab:limit} together with the observed upper limits on the number of events.

\begin{table}[!htbp]
\renewcommand{\arraystretch}{1.3}
 
\begin {center}
\caption{The observed and expected upper limits at 95\% confidence level on the visible cross section $\sigma\times A\times\epsilon$ from all signal regions.
The observed limits on the number of events are also reported as well as the fiducial efficiencies, $\epsilon$.}
\begin{tabular}{lcccc}
\toprule
\toprule
Signal region                        &  ($\sigma\times A\times\epsilon)_{\mathrm {obs}}^{95}$ [fb]  & ($\sigma\times A\times\epsilon)_{\mathrm   {exp}}^{95}$ [fb] & $N_{\mathrm  { obs}}^{95}$  & $\epsilon$ [\%] \\
\hline
 
SRI1&$2.45 $&$ 2.82^{+1.08}_{-0.78}$&$ 340$&$76$ \\
SRI2&$1.42 $&$ 1.68^{+0.63}_{-0.46}$&$ 198$&$74$ \\
SRI3&$0.93 $&$ 1.07^{+0.40}_{-0.29}$&$ 129$&$72$ \\
SRI4&$0.53 $&$ 0.63^{+0.23}_{-0.17}$&$ 74$&$67$ \\
SRE1&$1.80 $&$ 2.03^{+0.77}_{-0.56}$&$ 250$&$75$ \\
SRE2&$1.04 $&$ 1.15^{+0.43}_{-0.31}$&$ 145$&$75$ \\
SRE3&$0.79 $&$ 0.82^{+0.31}_{-0.22}$&$ 109$&$71$ \\

\bottomrule
\bottomrule
\end{tabular}
\label{tab:limit}
\end {center}
\end{table}

In order to provide additional constraints on BSM physics, which can be reinterpreted in terms of the results from this paper, a fiducial region is defined at the particle level with the same selection criteria as in the SRs.
The \metv\ computation at particle level is given by the vector sum of the transverse momenta of all non-interacting particles.
The $\met$ significance requirement is not applied, because  the object-based definition is not reproducible at particle level, but the impact of this selection on signal events is treated as negligible.
 
Given the fiducial acceptance ($A$) for a particular model and the fiducial efficiency ($\epsilon$), calculated as the ratio of the number of events passing the signal region selection at reconstruction level to the number of events passing the fiducial selection at the particle level, it is straightforward  to convert the visible cross-section limit into fiducial cross-section ($\sigma\times A$)  limits.
Using  the DM simplified model, described in Section~\ref{sec:MCSimulations}, the fiducial efficiency for each SR is calculated and the lowest values, shown in Table \ref{tab:limit}, can be used to set the fiducial cross-section limit in a conservative way.

Exclusion limits for the BSM models studied are obtained from the  `simplified shape fit' described in Section~\ref{sec:BGEstimation}.
A fit of the background plus the signal model is performed, where the signal component is allowed to populate both the SR and CRs,  with the signal strength being the freely floating signal normalisation factor. A specific signal is excluded if the upper limit on the signal strength is less than unity.

Systematic uncertainties in the signal predictions from the simplified DM and ALP models, described in Section~\ref{sec:MCSimulations}, are included in the fit.
The sources of experimental uncertainties are related to the physics objects as described for the main SM backgrounds.
Theoretical uncertainties include uncertainties in the NLO cross section due to  QCD factorisation and renormalisation scales~\cite{Bothmann:2016nao} and the choice of parton distribution functions. These uncertainties are below 5\% for acceptance and cross section, for different DM signals. Uncertainties in initial- and final-state radiation due to the choice of parton shower parameters used with \PYTHIA 8.2 are estimated by generating MC samples with the alternative tunes described in Ref.~\cite{ATL-PHYS-PUB-2014-021}, and are less than 10\%. The same sources of uncertainty are considered for ALP signals and the largest uncertainty is approximately  20\%, which comes mainly from QCD factorisation and renormalisation scales.

The results are presented for simplified DM models, described in Section~\ref{sec:MCSimulations}, with the exchange of an axial-vector or a vector mediator in the $s$-channel for  different couplings to quarks and leptons:   $g_q = 0.25$,  $g_\ell = 0$ and $g_q = 0.1$, $g_\ell = 0.1$ (0.01)~\cite{DMwg}.
 
As it was verified that the choice of mediator and couplings only affects the cross section and not the acceptance of the signal, the cross-section predictions for an axial-vector mediator  are rescaled in order to obtain the results for a vector mediator and for different couplings.
Observed and expected 95\% CL exclusion contours in the $m_{\chi}$--$m_{\textrm{med}}$   plane are shown in Figure~\ref{fig:DM_limits}.
The region of the plane under the limit curves is excluded.
The band around the expected contour shows the $\pm 1\sigma$  variations including all uncertainties described in Section~\ref{sec:results} except theoretical uncertainties affecting the signal cross section.
Those uncertainties are instead indicated as dotted lines around the observed limit.
The line corresponding to the DM thermal relic abundance measured by the Planck Collaboration~\cite{Planck2015} is also indicated~\cite{DMwg}.
The region not allowed due to perturbative unitarity violation is to the left of the line defined by  $m_{\chi} = \sqrt{\pi/2}m_{\textrm{med}}$~\cite{Perturb}.
The results of the search are summarised in  Table \ref{tab:DM_results}, where the values for $m_{\textrm{med}}$ and   $m_{\chi}$ correspond to the maximum excluded
values of the mediator masses and DM masses. These limits are more stringent than the limits obtained from the previous search using 36.1~\ifb~\cite{paper_monophoton2017},  which excluded dark-matter candidates with masses up to 340(480) GeV for axial-vector (vector) mediators, while the maximum excluded mass of the mediator was 1200 GeV for the same DM models.

\begin{table}[hbtp!]
\center
\caption{Observed limits at $95\%$ CL on $m_{\textrm{med}}$ and $m_{\chi}$ for the mediators and couplings to quarks, DM particles, and leptons considered for each model.
The values reported for $m_{\textrm{med}}$ and $m_{\chi}$ correspond to the maximum excluded values of mediator masses and DM masses in the search.}
\label{tab:DM_results}
\begin{tabular}{  lccccc }
\hline
\hline
Mediator    &    ~~~~$g_{q}$    &  ~~~~$g_{\chi} $    &     ~~~~$g_{\ell}$         & $m_{\textrm {med}}$ [\GeV]  & $m_{\chi}$ [\GeV]  \\
\hline
Axial-vector          &       ~~~~0.25       &     ~~1                &   0   &  1460 & 415 \\
Axial-vector         &          ~~0.1        &      ~~1          &   ~~~0.1   &  ~~920 &  280\\
Vector     &        ~~~~0.25      &      ~~1                &    0              & 1470  &  580 \\
Vector     &           ~~0.1    &      ~~1               &    ~~~~~0.01        & ~~950 & 400    \\

\hline
\hline
\end{tabular}
\end{table}

To show the complementarity with DM direct detection searches, the contours in the $m_{\chi}$--$m_{\textrm{med}}$ plane obtained for a specific choice of mediator and couplings can be directly translated into bounds on the $\chi$--nucleon scattering cross section following the procedure described  in Ref.~\cite{DM_DD}.
Figure~\ref{fig:DM_DD_limits} (top left) shows the 90\% CL exclusion limits on the $\chi$--proton spin-dependent (SD) scattering cross section versus $m_{\chi}$ in the axial-vector model with couplings $g_{q}$ = 0.25, $g_{\chi}$ = 1 and $g_{\ell}$ = 0. In Figure~\ref{fig:DM_DD_limits} (top right), 90\% CL exclusion limits on the $\chi$--neutron SD scattering cross section versus $m_{\chi}$ are shown for the same model. A comparison with the results from direct DM searches~\cite{Picasso, XENON1T, CDM, LUX, PICO, PANDA} is also shown. The search probes complementary regions with respect to direct DM searches in the full parameter space, providing greater sensitivity for $m_{\chi}$ values below 440~\GeV\ for these models and parameter values.
Figure~\ref{fig:DM_DD_limits} (bottom) shows the 90\% CL exclusion limits on the $\chi$--nucleon spin-independent (SI) scattering cross section versus $m_{\chi}$ for the vector model with couplings $g_{q}$ = 0.25, $g_{\chi}$ = 1 and $g_{\ell}$ = 0.  With the exception of small $m_{\chi}$ masses, lower than about 3~\GeV, direct detection searches~\cite{XENON_ID, MIGD, PandaX_ID, DarkSide_ID} provide stronger limits for these models and parameter values.

The results are also interpreted in terms of limits on the parameters of the ALP model described in Section~\ref{sec:MCSimulations}, obtained for the first time using the full Run-2 dataset from events with a $\gam + \met$ final state.
Figure~\ref{fig:alps_cwtil_vs_fa_wZinvLimit}  shows the expected and observed limits at $95\%$ CL on the coefficient $\CW$ as a function of the effective scale $f_a$ for an ALP mass of 1~\MeV.
The limits are obtained by using the cross section to rescale the result obtained for the point generated with $\CW = 1$, $f_a = 1$~\TeV\ and $m_a = 1$~\MeV.  The limits on $\CW$ increase linearly with $f_a$. 
For $f_a = 1000$~\GeV\ the couplings $\CW > 0.12\,(0.13)$ are excluded according to the observed (expected) limit. The ALP model is an EFT and it becomes invalid for $\hat{s}$ > $f_a^2$ ($\hat{s}$ corresponds to the invariant mass-squared of the partonic collision).
The validity of the limit is verified by applying a suppression factor $f_a^4/\hat{s}^2$ to the events outside of the region of validity; this truncation has an almost negligible impact for $f_a > 1000$~\GeV.
The largest change in the limit is 1\% for $f_a$ = 1~\TeV, where the truncation impact is largest.
 
Since the signal cross section  and width depend on the ratio of the operator coefficient to the effective scale, the limit is also computed for  $\CW$/$f_a$  as a function of ALP mass in the range from 1~\MeV\ to 1~\GeV, where the acceptance is constant. The result obtained for the observed (expected) upper limit is   $\CW/f_a =1.2 \times10^{-4}$ $(1.3 \times 10^{-4})$~\GeV$^{-1}$,
constant with the ALP mass in the considered range. This limit constrains  the coupling  of the ALP to the electroweak gauge bosons, $|g_{aZ\gamma}| < 0.51~\TeV^{-1}$ assuming $g_{a\gamma\gamma}=0$.  In addition, using the relationship ~\cite{Brivio:2017ije}:
\begin{equation}\nonumber
\Gamma(Z\to a \gamma) = \frac{m_Z^3}{384\pi} g_{aZ\gamma}^2 \left(1 - \frac{m_a^2}{m_Z^2}\right)^3,
\end{equation}
and assuming   $m_a =  1$~\MeV, a contribution  to the $Z$ boson width of  $\Gamma(Z\to a \gamma) < 0.17$~\MeV\  at 95\% CL is estimated.
The most stringent limit is set from LEP,  $\Gamma(Z\to \textrm{X}\gamma) < 2.5$~\keV\ for photons with energy above $30~\GeV$ in $ e^+ e^-$ annihilation events at the $Z$ resonance  can be inferred from Ref.~\cite{Acciarri:1997im}, where X refers to a stable, weakly interacting particle.

\begin{figure}[!htbp]\centering
\includegraphics[width=7.5cm,height=7cm]{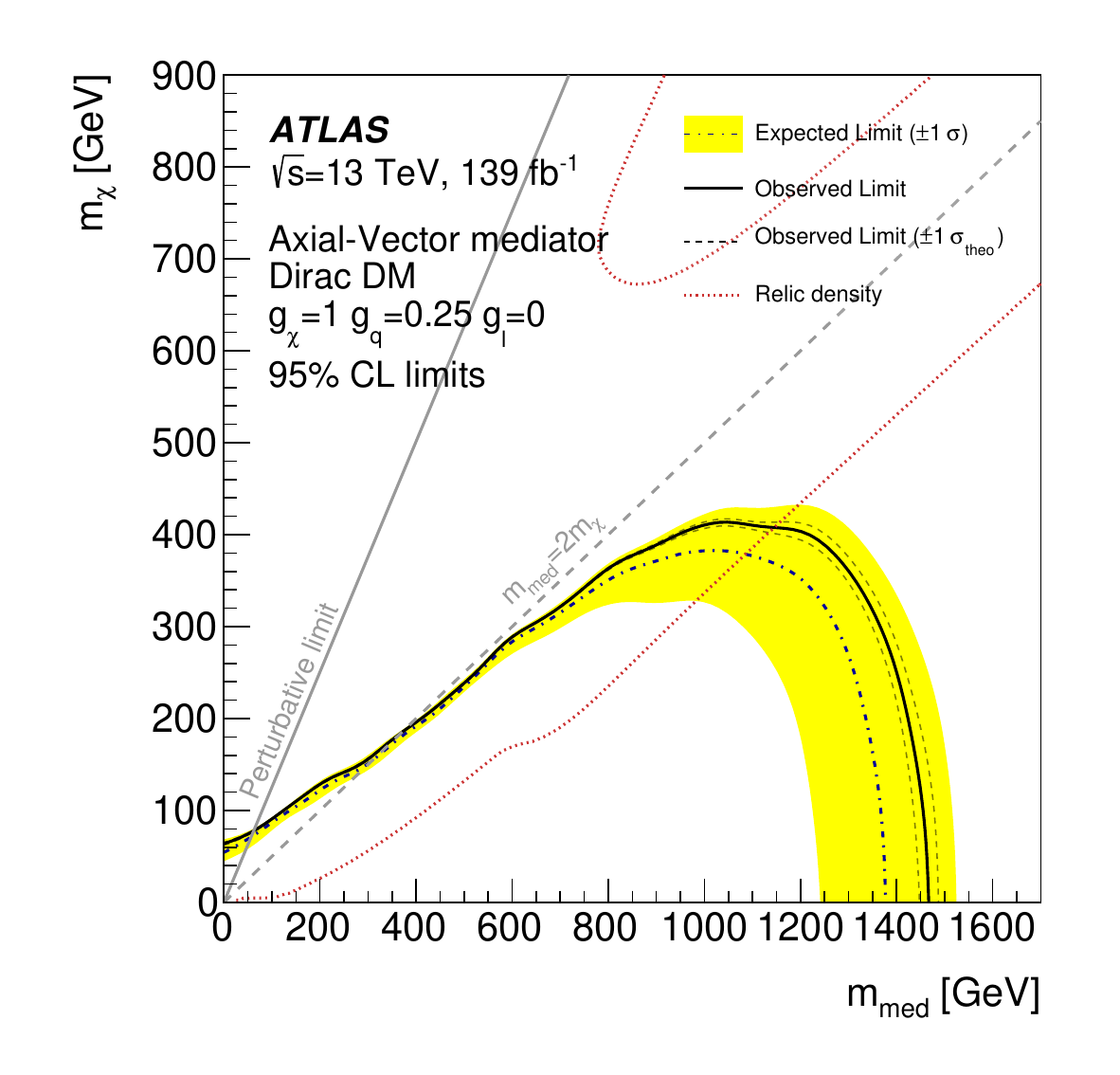} \includegraphics[width=7.5cm,height=7cm]{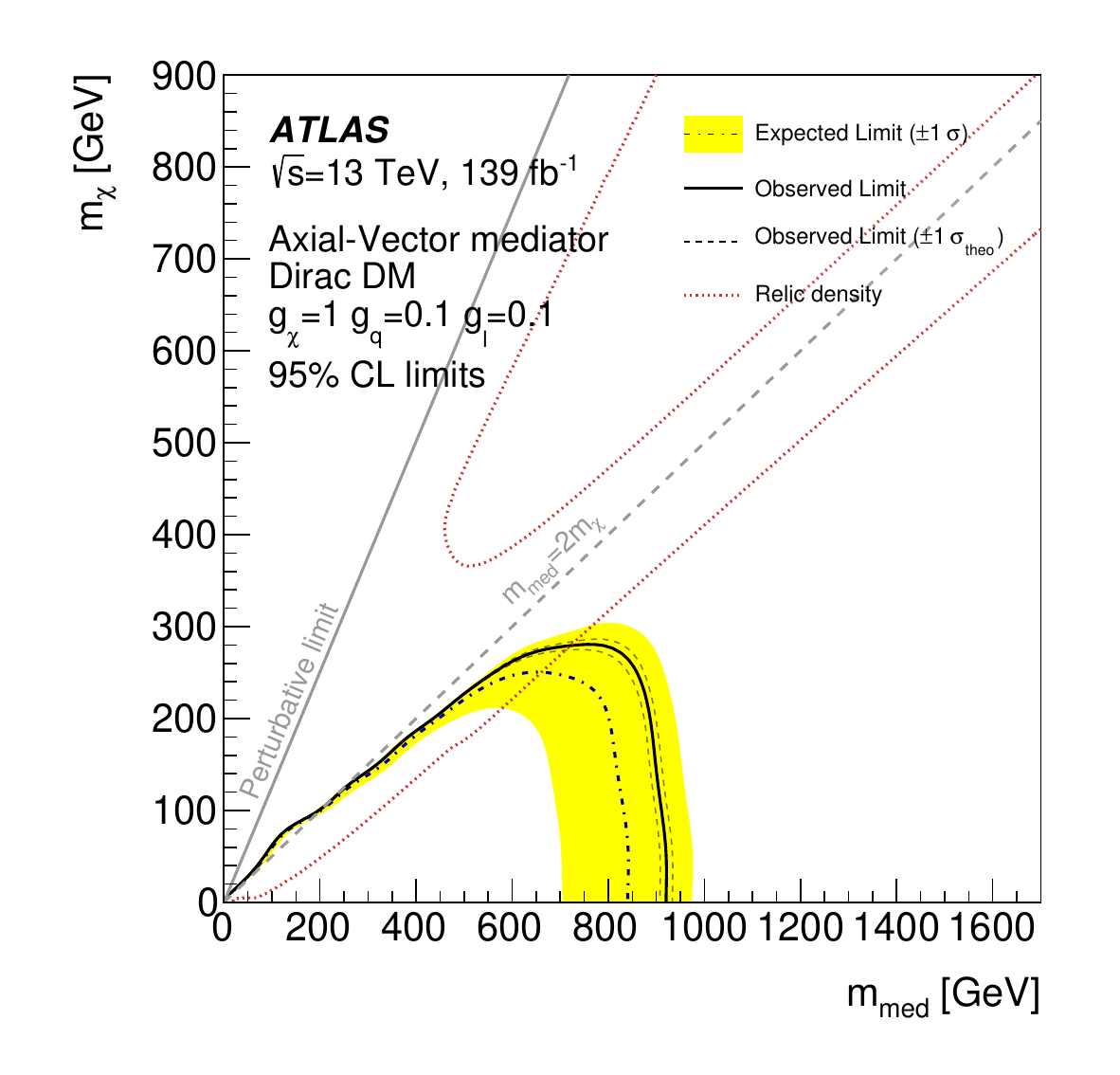} \\
\includegraphics[width=7.5cm,height=7cm]{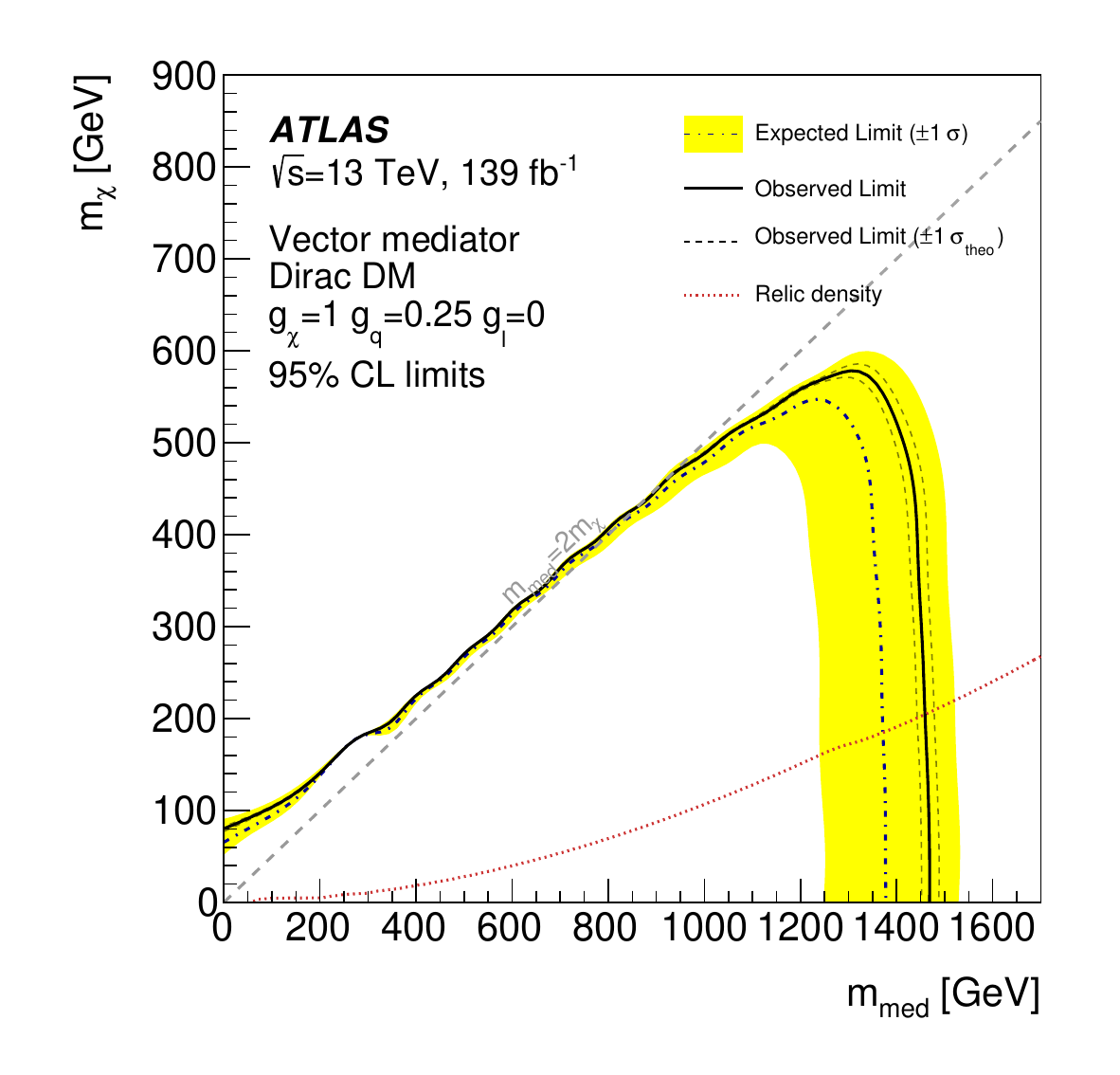} \includegraphics[width=7.5cm,height=7cm]{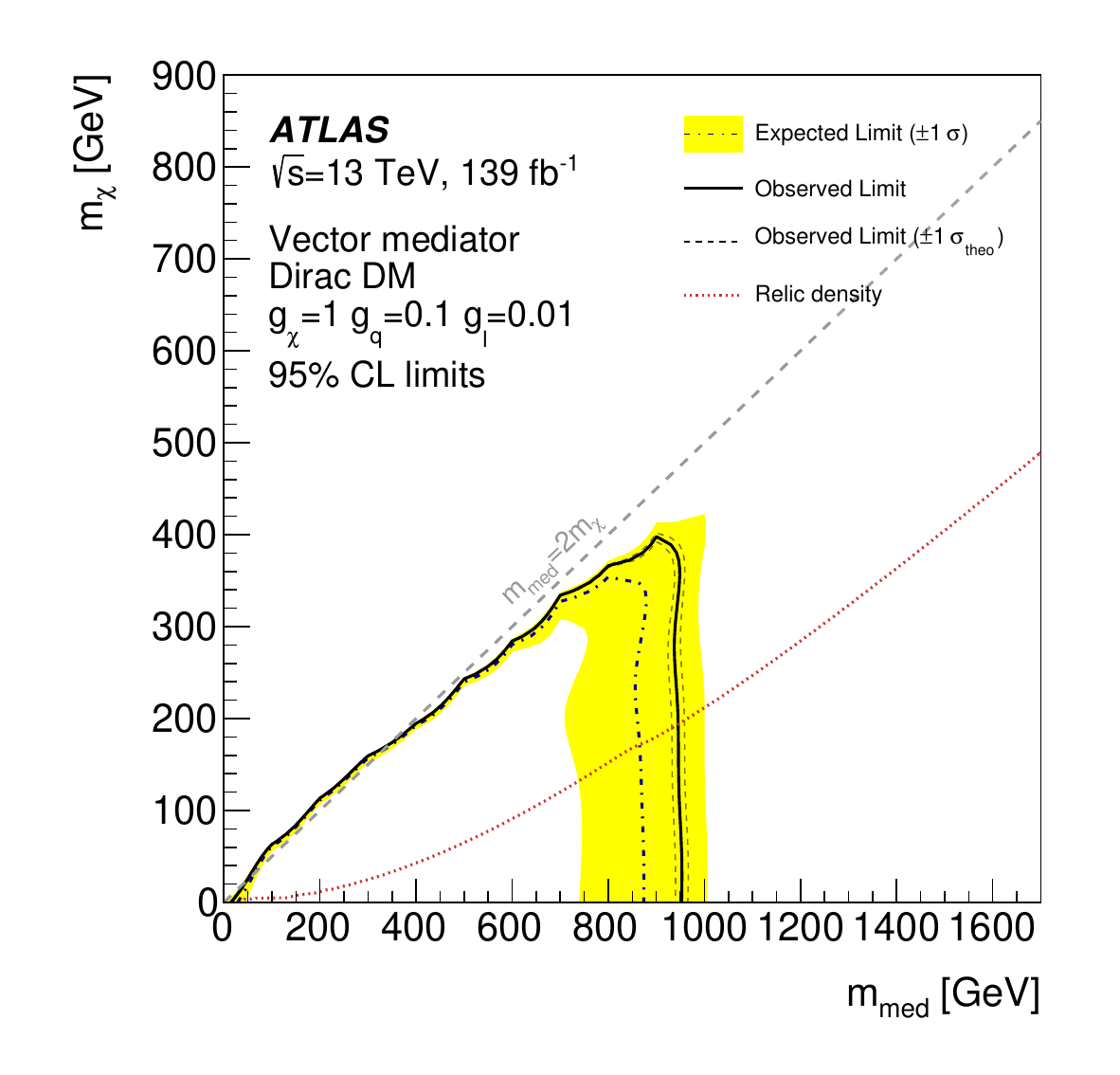}
 
\caption{The observed (solid line) and expected (dot-dashed line) 95\% CL exclusion contours  in the $m_{\chi}$--$m_{\textrm{med}}$ plane for a simplified DM model involving an axial-vector mediator with  couplings $g_{\chi}$ = 1, $g_q$ = 0.25 and $g_\ell$ = 0 (top left) and  $g_{\chi}$ = 1, $g_q$ = 0.1 and  $g_\ell$ = 0.1 (top right).
The same is shown for a vector mediator  with couplings $g_{\chi}$ = 1, $g_q$ = 0.25 and $g_\ell$ = 0 (bottom left) and $g_{\chi}$ = 1, $g_q$ = 0.1 and  $g_\ell$ = 0.01 (bottom right).
The area under the limit curve is excluded. The region to the left of the line defined by  $m_{\chi} = \sqrt{\pi/2}m_{\textrm{med}}$ is excluded by the perturbative limit which is relevant for axial-vector mediators. The relic density curve  ~\cite{Planck2015} is also shown.
The area  below the relic density curve in the on-shell region (or above in the off-shell region in the axial-vector mediator case)  corresponds to a predicted DM overabundance. }
 
\label{fig:DM_limits}
\end{figure}

\begin{figure}[!htbp]\centering
\includegraphics[width=7.5cm,height=7cm]{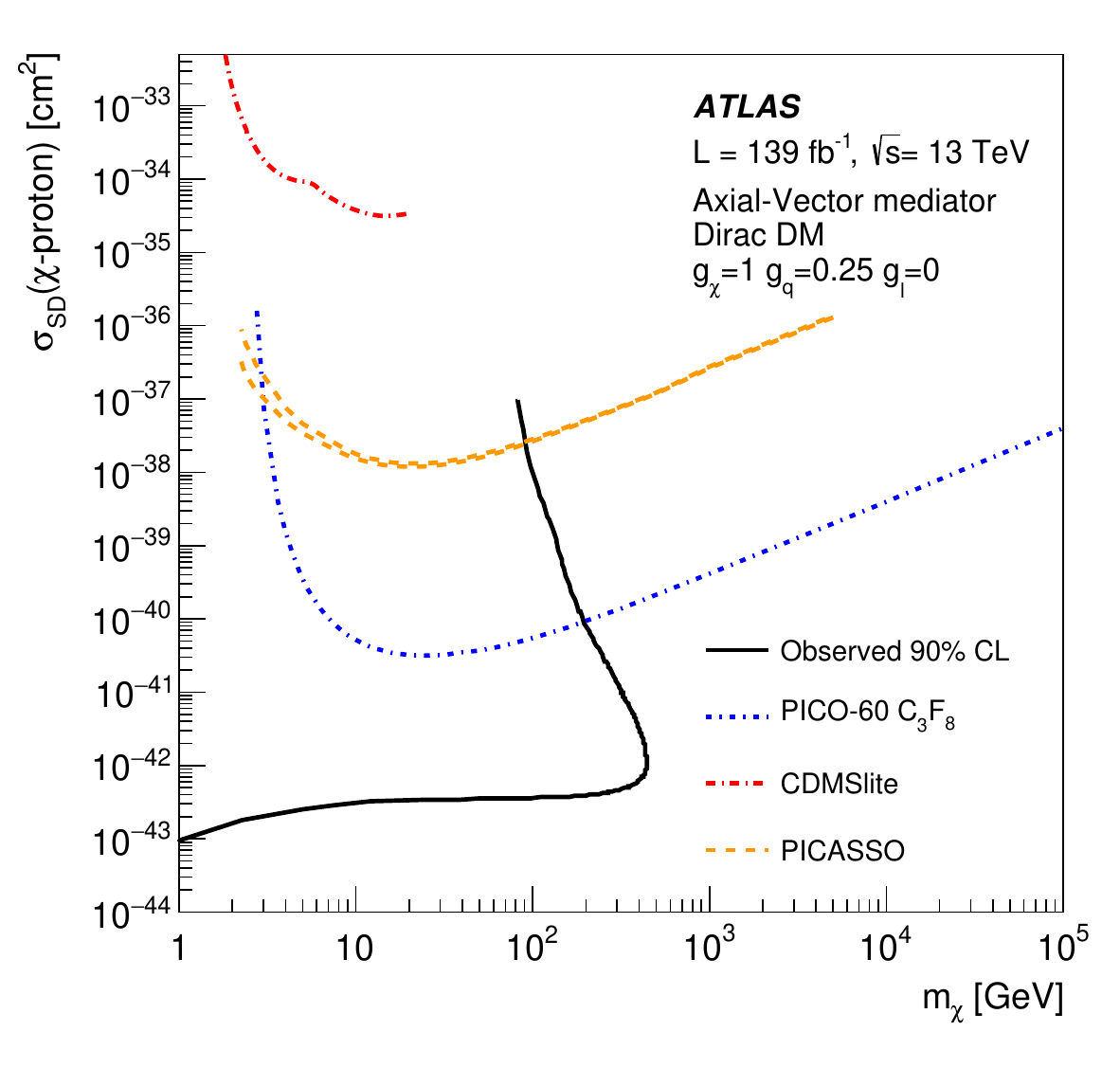} \includegraphics[width=7.5cm,height=7cm]{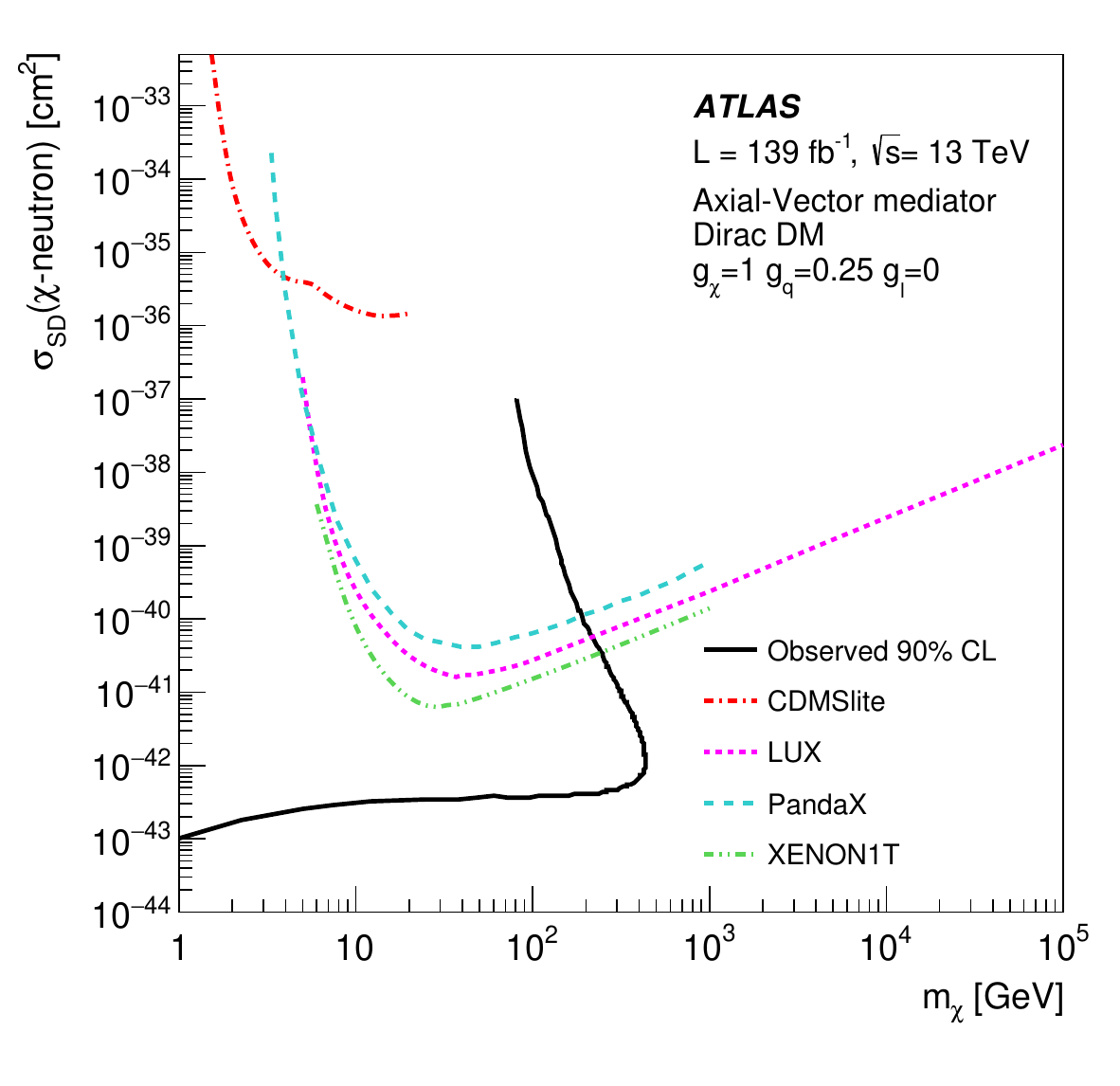} \\
\includegraphics[width=7.5cm,height=7cm]{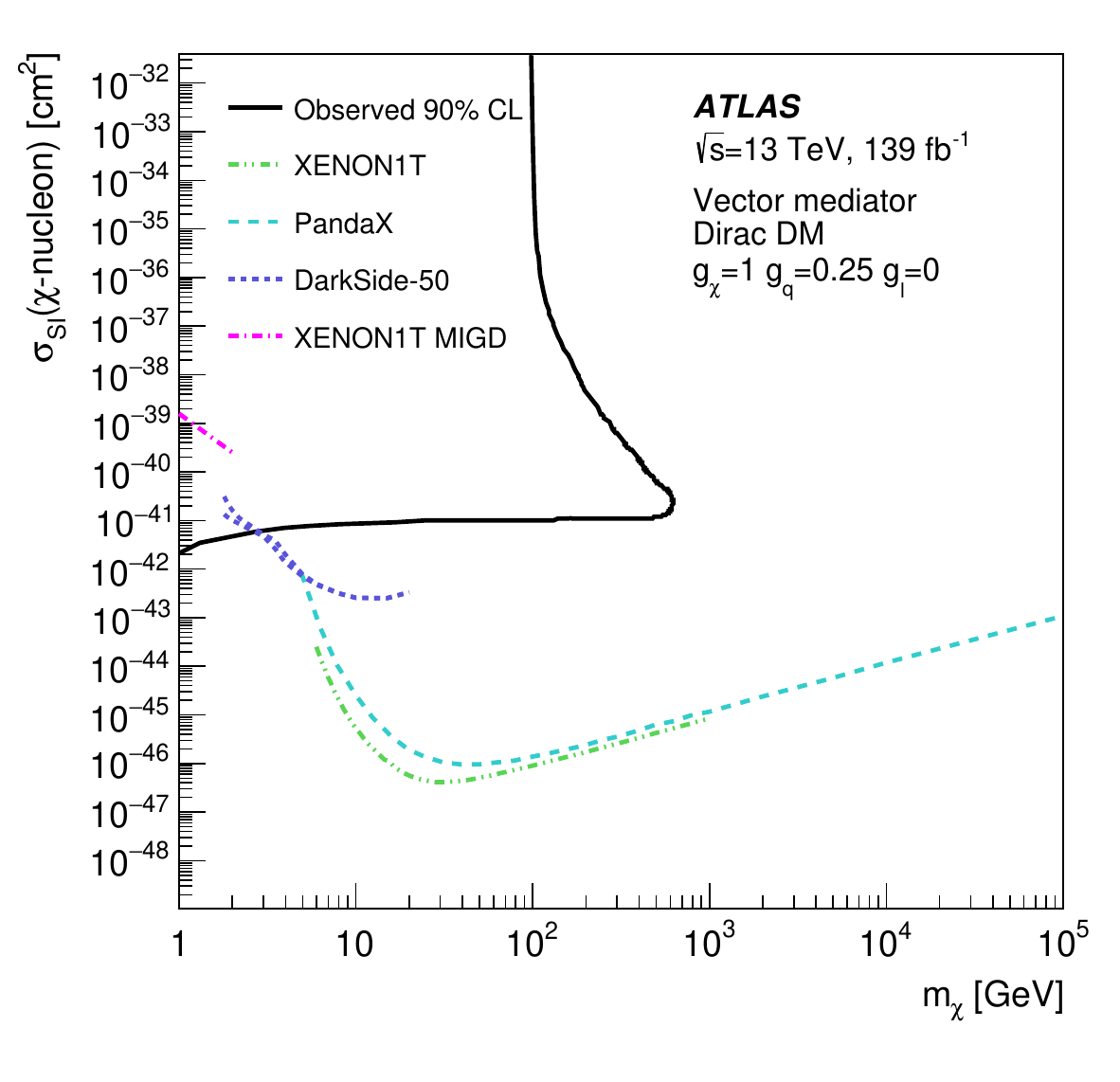}
\caption{The 90\% CL exclusion limit on the $\chi$--proton  spin-dependent scattering cross section (top left) and on the $\chi$--neutron spin-dependent scattering cross section (top right) in an axial-vector model with couplings  $g_{\chi} = 1$, $g_q = 0.25$ and $g_\ell = 0$  as a function of the dark-matter mass $m_{\chi}$.  The 90\% CL exclusion limit  is displayed only in the regime where the theory is perturbative. Results at  90\% CL from direct DM searches~\cite{Picasso, XENON1T, CDM, LUX, PICO, PANDA} are also shown.
The 90\% CL exclusion limit on the $\chi$--nucleon spin-independent  scattering cross section (bottom) in a vector model with couplings  $g_{\chi} = 1$, $g_q = 0.25$ and $g_\ell = 0$  as a function of the dark-matter mass $m_{\chi}$. Results at  90\% CL from direct DM searches~\cite{XENON_ID, MIGD, PandaX_ID, DarkSide_ID} are also shown.
}
\label{fig:DM_DD_limits}
\end{figure}

\begin{figure}[ht!]
\centering
\includegraphics[width=7.5cm,height=7cm]{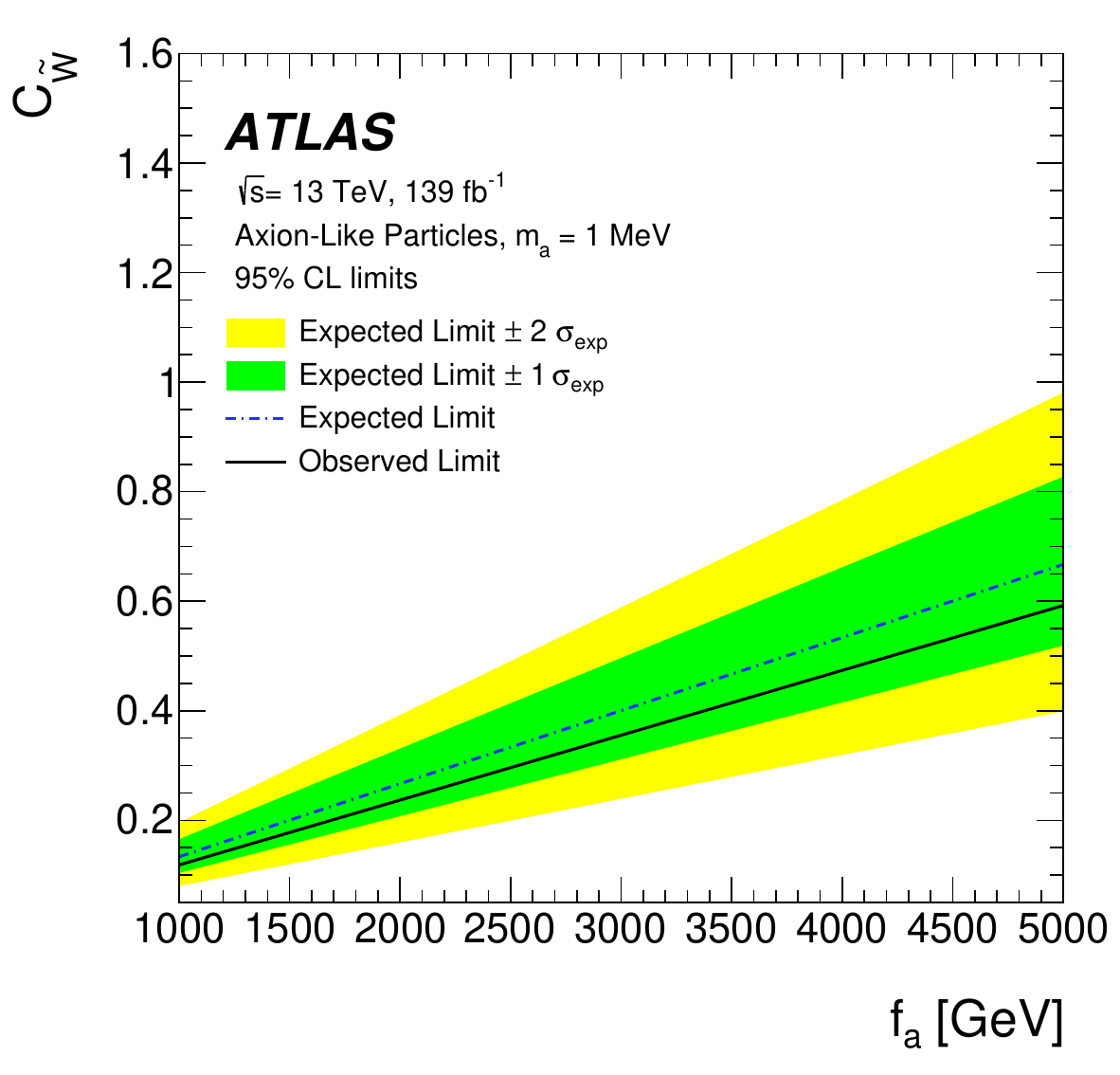}
\caption{Observed (solid line) and expected (dot-dashed line) exclusions at $95\%$ CL  on the coupling  $\CW$ as a function of the effective scale $f_a$  for an  ALP mass of 1~\MeV. The region above the limit lines is excluded.}
\label{fig:alps_cwtil_vs_fa_wZinvLimit}
\end{figure}

 
\clearpage
\section{Conclusion}
 
A search for an excess of events in  a $\gam + \met$ final state over the SM prediction is performed using  139~\ifb\  of proton--proton data collected by the ATLAS experiment at the LHC at a centre-of-mass energy of $\sqrt{s} = 13~\TeV$.
Events with an isolated photon with transverse energy above 150~\GeV\ in association with $\met > 200~\GeV$ are selected. Several signal regions  (inclusive and exclusive) are defined for different \met\ ranges to improve the sensitivity, and control regions are adopted to estimate the main backgrounds, decreasing the impact of the experimental and theoretical uncertainties with a data-driven approach.
The observed data are consistent with the SM expectation. 
Model-independent 95\% CL upper limits are set on the visible cross section for events beyond the Standard Model ranging from 2.45 to 0.5~\fb\ in signal regions corresponding to different \met\ ranges.
Model-dependent 95\% CL limits are placed on parameters of simplified dark-matter models.
Dark-matter candidates are excluded for masses up to 415 (580) GeV for axial-vector (vector) mediators, while the maximum excluded mass of an axial-vector (vector) mediator is 1460 (1470)~\GeV.
The results are also translated into limits on the parameters of the axion-like particle (ALP) model. The $95\%$ CL limits on the coupling $\CW$  are computed as a function of the effective scale $f_a$  for an ALP mass of 1~\mev.  For $f_a = 1$~\TeV, values of the couplings $\CW > 0.12\,(0.13)$ are excluded according to the observed (expected) limit. The observed limit constrains the coupling  of the ALP to the electroweak gauge bosons to be $|g_{aZ\gamma}| < 0.51~\TeV^{-1}$, assuming $g_{a\gamma\gamma}=0$.
 
\clearpage
 
\begin{flushleft}
\hypersetup{urlcolor=black}
{\Large The ATLAS Collaboration}

\bigskip

\AtlasOrcid[0000-0002-6665-4934]{G.~Aad}$^\textrm{\scriptsize 102}$,    
\AtlasOrcid[0000-0002-5888-2734]{B.~Abbott}$^\textrm{\scriptsize 128}$,    
\AtlasOrcid[0000-0002-7248-3203]{D.C.~Abbott}$^\textrm{\scriptsize 103}$,    
\AtlasOrcid[0000-0002-2788-3822]{A.~Abed~Abud}$^\textrm{\scriptsize 36}$,    
\AtlasOrcid[0000-0002-1002-1652]{K.~Abeling}$^\textrm{\scriptsize 53}$,    
\AtlasOrcid[0000-0002-2987-4006]{D.K.~Abhayasinghe}$^\textrm{\scriptsize 94}$,    
\AtlasOrcid[0000-0002-8496-9294]{S.H.~Abidi}$^\textrm{\scriptsize 167}$,    
\AtlasOrcid[0000-0002-8279-9324]{O.S.~AbouZeid}$^\textrm{\scriptsize 40}$,    
\AtlasOrcid{N.L.~Abraham}$^\textrm{\scriptsize 156}$,    
\AtlasOrcid[0000-0001-5329-6640]{H.~Abramowicz}$^\textrm{\scriptsize 161}$
% ...
\bigskip
\\

% ...
$^{36}$CERN, Geneva; Switzerland.\\
% ...
$^{40}$Niels Bohr Institute, University of Copenhagen, Copenhagen; Denmark.\\
% ...
$^{53}$II. Physikalisches Institut, Georg-August-Universit\"{a}t G\"ottingen, G\"ottingen; Germany.\\
% ...
$^{94}$Department of Physics, Royal Holloway University of London, Egham; United Kingdom.\\
% ...
$^{102}$CPPM, Aix-Marseille Universit\'e, CNRS/IN2P3, Marseille; France.\\
$^{103}$Department of Physics, University of Massachusetts, Amherst MA; United States of America.\\
% ...
$^{128}$Homer L. Dodge Department of Physics and Astronomy, University of Oklahoma, Norman OK; United States of America.\\
% ...
$^{156}$Department of Physics and Astronomy, University of Sussex, Brighton; United Kingdom.\\
% ...
$^{161}$Raymond and Beverly Sackler School of Physics and Astronomy, Tel Aviv University, Tel Aviv; Israel.\\
% ...
$^{167}$Department of Physics, University of Toronto, Toronto ON; Canada.\\
% ...
$^{s}$ Also at Giresun University, Faculty of Engineering, Giresun; Turkey.\\
% ...

\end{flushleft}


\clearpage
\section*{Acknowledgements}
 

We thank CERN for the very successful operation of the LHC, as well as the
support staff from our institutions without whom ATLAS could not be
operated efficiently.
 
We acknowledge the support of ANPCyT, Argentina; YerPhI, Armenia; ARC, Australia; BMWFW and FWF, Austria; ANAS, Azerbaijan; SSTC, Belarus; CNPq and FAPESP, Brazil; NSERC, NRC and CFI, Canada; CERN; ANID, Chile; CAS, MOST and NSFC, China; COLCIENCIAS, Colombia; MSMT CR, MPO CR and VSC CR, Czech Republic; DNRF and DNSRC, Denmark; IN2P3-CNRS and CEA-DRF/IRFU, France; SRNSFG, Georgia; BMBF, HGF and MPG, Germany; GSRT, Greece; RGC and Hong Kong SAR, China; ISF and Benoziyo Center, Israel; INFN, Italy; MEXT and JSPS, Japan; CNRST, Morocco; NWO, Netherlands; RCN, Norway; MNiSW and NCN, Poland; FCT, Portugal; MNE/IFA, Romania; JINR; MES of Russia and NRC KI, Russian Federation; MESTD, Serbia; MSSR, Slovakia; ARRS and MIZ\v{S}, Slovenia; DST/NRF, South Africa; MICINN, Spain; SRC and Wallenberg Foundation, Sweden; SERI, SNSF and Cantons of Bern and Geneva, Switzerland; MOST, Taiwan; TAEK, Turkey; STFC, United Kingdom; DOE and NSF, United States of America. In addition, individual groups and members have received support from BCKDF, CANARIE, Compute Canada, CRC and IVADO, Canada; Beijing Municipal Science \& Technology Commission, China; COST, ERC, ERDF, Horizon 2020 and Marie Sk{\l}odowska-Curie Actions, European Union; Investissements d'Avenir Labex, Investissements d'Avenir Idex and ANR, France; DFG and AvH Foundation, Germany; Herakleitos, Thales and Aristeia programmes co-financed by EU-ESF and the Greek NSRF, Greece; BSF-NSF and GIF, Israel; La Caixa Banking Foundation, CERCA Programme Generalitat de Catalunya and PROMETEO and GenT Programmes Generalitat Valenciana, Spain; G\"{o}ran Gustafssons Stiftelse, Sweden; The Royal Society and Leverhulme Trust, United Kingdom.
 
The crucial computing support from all WLCG partners is acknowledged gratefully, in particular from CERN, the ATLAS Tier-1 facilities at TRIUMF (Canada), NDGF (Denmark, Norway, Sweden), CC-IN2P3 (France), KIT/GridKA (Germany), INFN-CNAF (Italy), NL-T1 (Netherlands), PIC (Spain), ASGC (Taiwan), RAL (UK) and BNL (USA), the Tier-2 facilities worldwide and large non-WLCG resource providers. Major contributors of computing resources are listed in Ref.~\cite{ATL-SOFT-PUB-2020-001}.
 

\printbibliography

\end{document}